\documentclass[twocolumn,showpacs,aps,pra,amsmath,amssymb,superscriptaddress]{revtex4-1}
\usepackage{bm,color,bbm}
\usepackage{ulem} 
\usepackage{hyperref, mathtools,graphicx,natbib}
\usepackage{subfigure}

\newcommand{\beq}{\begin{equation}}
\newcommand{\eeq}{\end{equation}}
\newcommand{\bqa}{\begin{eqnarray}}
\newcommand{\eqa}{\end{eqnarray}}
\newcommand{\nn}{\nonumber}

\newcommand{\smallfrac}[2]{\mbox{$\frac{#1}{#2}$}}

\newcommand{\half}{\smallfrac{1}{2}}

\definecolor{maroon}{rgb}{0.7,0,0}

\definecolor{ngreen}{rgb}{0.3,0.7,0.3}

\definecolor{golden}{rgb}{0.8,0.6,0.1}

\begin{document}
\title{{Measurement-dependence cost for Bell nonlocality: causal vs retrocausal models}
}
\author{Michael J. W. Hall}
\affiliation{Department of Theoretical Physics, Research School of Physics, Australian National University, Canberra ACT 0200, Australia}
\author{Cyril Branciard}
\affiliation{Universit\'e Grenoble Alpes, CNRS, Grenoble INP, Institut N\'eel, 38000 Grenoble, France}

\date{\today}

\begin{abstract}
Device independent protocols based on Bell nonlocality, such as quantum key distribution and randomness generation, must ensure no adversary can have prior knowledge of the measurement outcomes. This requires a measurement independence assumption: that the choice of measurement is uncorrelated with any other underlying variables that influence the measurement outcomes. Conversely, relaxing measurement independence allows for a fully `causal' simulation of Bell nonlocality. We construct the most efficient such simulation, as measured by the mutual information between the underlying variables and the measurement settings, for the Clauser-Horne-Shimony-Holt (CHSH) scenario, and find that the maximal quantum violation requires a mutual information of  just $\sim 0.080$ bits. Any physical device built to implement this simulation allows an adversary to have full knowledge of a cryptographic key or `random' numbers generated by a device independent protocol based on violation of the CHSH inequality. We also show that a previous model for the CHSH scenario, requiring only $\sim 0.046$ bits to simulate the maximal quantum violation, corresponds to the most efficient `retrocausal' simulation, in which future measurement settings necessarily influence earlier source variables. This may be viewed either as an unphysical limitation of the prior model, or as an argument for retrocausality on the grounds of its greater efficiency. Causal and retrocausal models are also discussed for maximally entangled two-qubit states, as well as superdeterministic, one-sided and zigzag causal models.
\end{abstract}

\maketitle

\section{Introduction}
\label{sec:intro}

Quantum information protocols that promise the secure distribution of cryptographic keys, or the generation of guaranteed randomness, must rely on the validity of certain physical assumptions. The strongest protocols, in the sense of requiring the weakest assumptions, are both device and theory independent---they have no reliance on internal details of preparation and measuremement devices, nor even on whether quantum mechanics is valid. Instead, their promise is based on witnessing the phenomenon of Bell nonlocality, independently of the physical means by which it is generated~\cite{bellreview, ekert2014}.

In particular, if some set of statistical correlations between spacelike separated observers violates a Bell inequality~\cite{bell1964}, then any model of the correlations must either (i) permit superluminal influences between the observation regions, or (ii) include underlying variables that  influence both the measurement outcomes and the choice of measurements, or (iii) have intrinsically random and unpredictable measurement outcomes~\cite{bellreview}. Thus, if one rules out the first two options by assumption, then intrinsic randomness is guaranteed. There can, therefore, be no eavesdropper or adversary holding a list of predetermined measurement outcomes, leading to the promised security of  protocols based on such correlations~\cite{bellreview, ekert2014}.

It follows that the assumptions ruling out options~(i) and~(ii) above require very careful scrutiny. Can observers who wish to make secure transactions trust the assertion of a device manufacturer that there are no  (possibly hidden) superluminal influences, nor any subtle influences on the choice of measurements (`measurement independence')---particularly if these choices are determined by random number generators supplied by the manufacturer? Or is there a loophole that an adversary can exploit to generate lists of predetermined measurement outcomes?

In  regard to no superluminal influences the observers can be easily reassured:  it is simply not known how to build classical devices that can influence each other superluminally (even if this might one day be possible, e.g., via some new wormhole technology). However, the case of measurement independence is not so reassuring: it is certainly possible to build devices that exhibit Bell nonlocality by violating this assumption, with no superluminal effects and with all measurement outcomes being predetermined~\cite{dialectica,brans,kofler2006,koh2012}. 

Surprisingly, however, not all measurement-dependent models of Bell nonlocality can be implemented `causally'---where by `causal' we mean here that it should be the underlying variables (which we consider to originate from a source in the past) that influence the (future) choice of measurement settings, rather than the other way around. Indeed, we will show here that the most efficient models can only be implemented `retrocausally', with future measurement selections influencing past source variables. Thus, similarly to superluminal models, they cannot be implemented with known technology. We are therefore led to seek models suitable for implementing Bell nonlocality in a fully causal manner, and to find the minimum information-theoretic resources that they require. An adversary with such resources can then build physical devices that subvert device independent protocols. 

\begin{figure*}[!ht]
	\centering
	\subfigure[~Causal measurement dependence]{
		\includegraphics[width=0.45\textwidth]{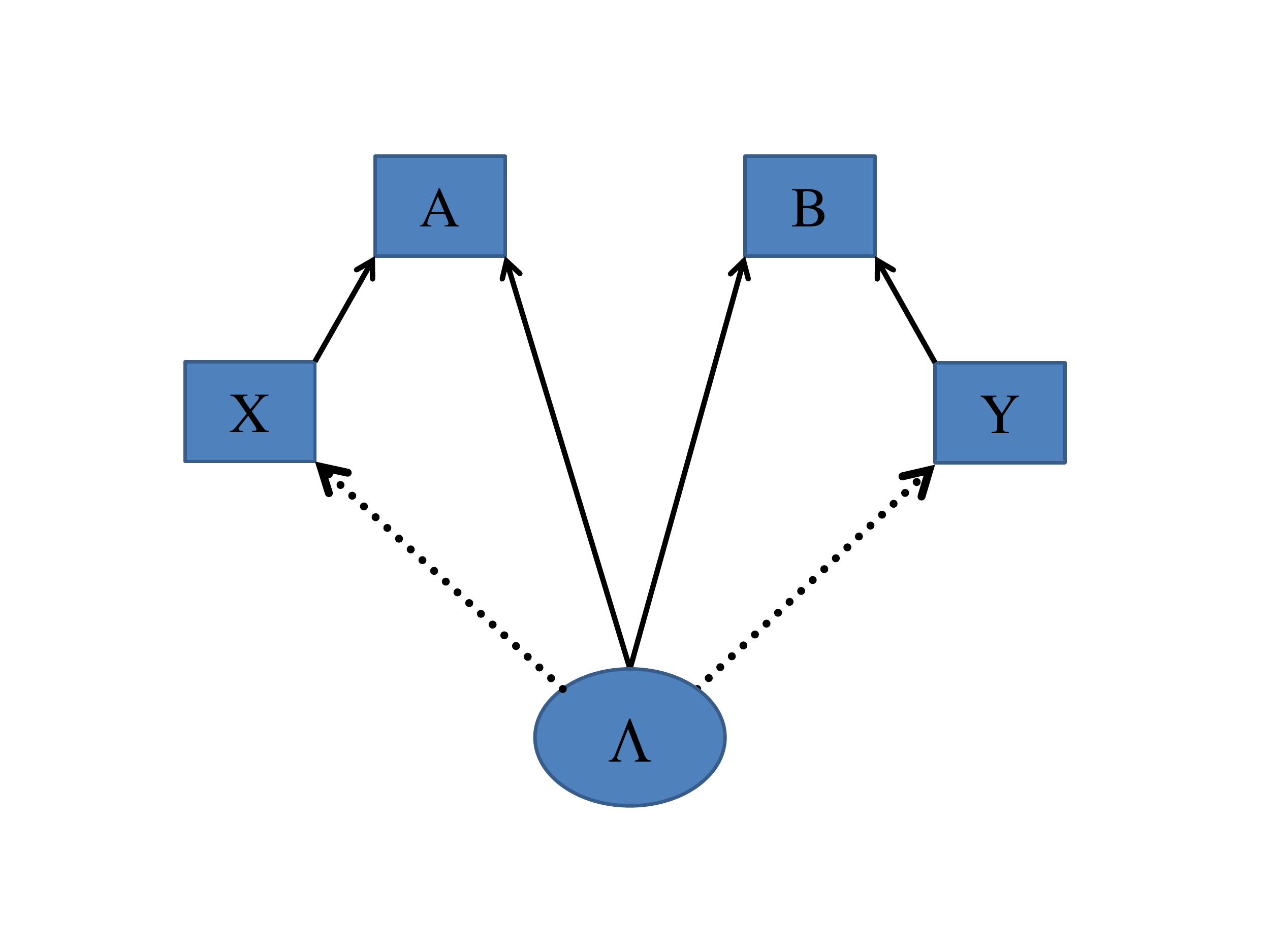}
		\label{fig:causal}
	}
	\subfigure[~Retrocausal measurement dependence]{
		\includegraphics[width=0.45\textwidth]{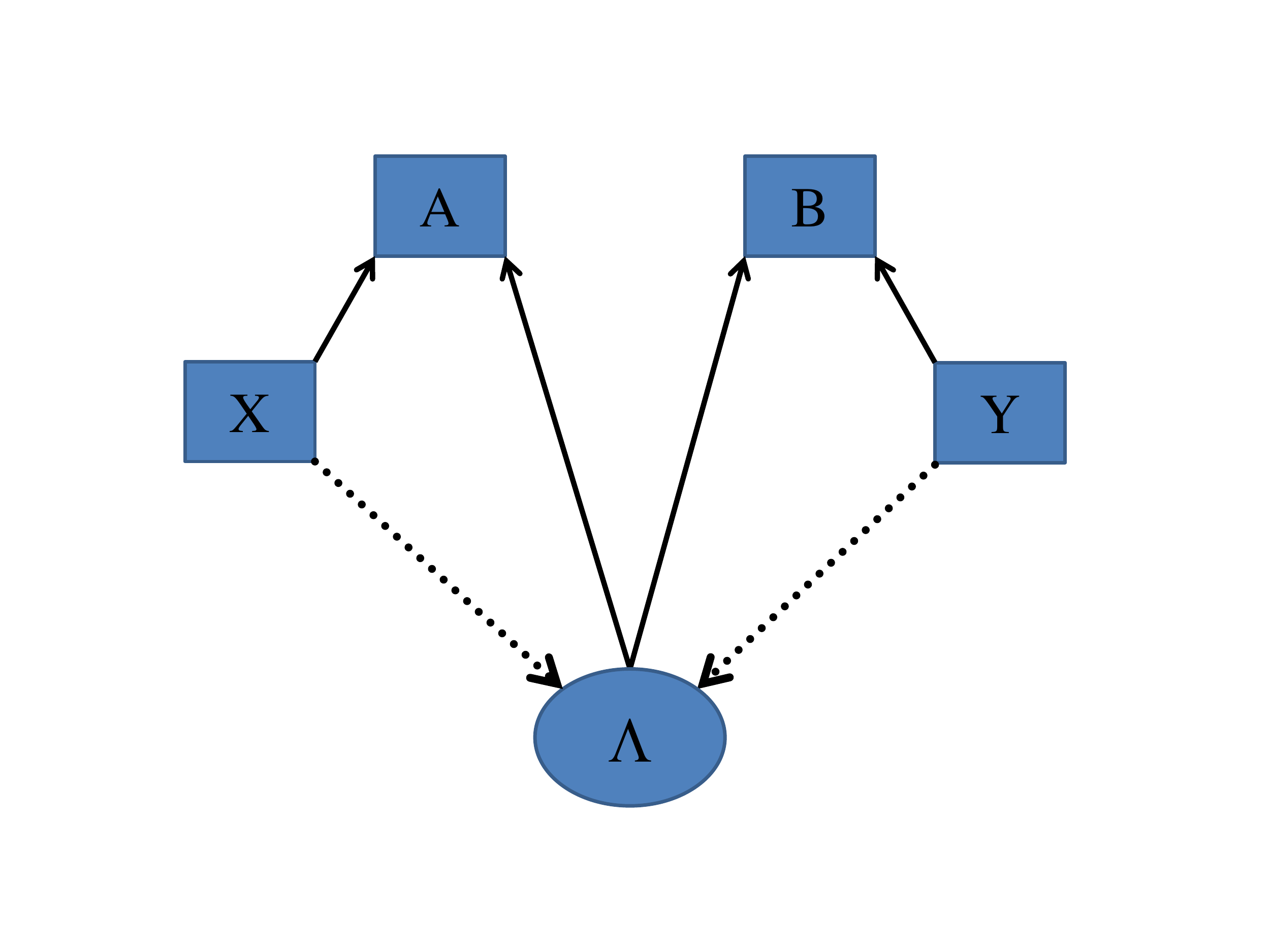}
		\label{fig:retro}
	}
	\caption{Examples of causal structures, for separable models of Bell nonlocality as per Eq.~(\ref{sep}). Each box is labelled by a corresponding random variable, with arrows indicating allowed causal influences between the boxes.  Thus, the values of a source variable $\Lambda$ and measurement selection variable $X$ ($Y$) can causally influence measurement outcome $A$ ($B$) (solid  arrows). Two types of measurement dependence are indicated by the dotted arrows. (a) Causal measurement dependence: the source variable can influence the measurement selections. (b) Retrocausal measurement dependence: the source variable can be influenced by the measurement selections.  Note that while causal structures do not inherently have a direction of time~\cite{pearl}, it is convenient here (particularly when considering practical implementations of causal structures), to regard them as being embedded in a relativistic spacetime, so that the upwards and horizontal directions in the figure are timelike and spacelike, respectively. This allows `causal', 'retrocausal' and `superluminal' causal influences to be unambiguously distinguished. Other examples are given in Sec.~\ref{sec:chsh}.
	}
	\label{fig:causvsretro}
\end{figure*}

In Sec.~\ref{sec:sep} we review models for Bell nonlocal correlations between two observers, the quantification of measurement dependence via mutual information, and the fundamental distinction between causal and retrocausal models of measurement dependence (see Fig.~\ref{fig:causvsretro}). In Sec.~\ref{sec:chsh} we specialise to the Clauser-Horne-Shimony-Holt (CHSH) scenario, in which each observer has two unbiased measurement choices with binary  outcomes~\cite{chsh,bellreview}. We determine measurement dependent models that require the least possible mutual information to model a given arbitrary violation of the CHSH Bell inequality, for several causal structures of interest. These models have deterministic measurement outcomes and no superluminal influences. The corresponding minimal informations are found to increase progressively for retrocausal, causal, one-sided, zigzag and superdeterministic measurement dependence as we consider here. In particular, retrocausal measurement dependence requires strictly less mutual information than any other type. The results also verify a recent optimality conjecture made in Ref.~\cite{fried2019}.

In Sec.~\ref{sec:biased} we show, by explicit construction, that biased measurement choices in the CHSH scenario reduce the minimum amount of mutual information required, in comparison to the unbiased case, for each of the causal structures considered. Hence the mutual informations calculated in Sec.~\ref{sec:chsh} are sufficient to simulate CHSH Bell nonlocality irrespective of the probabilities with which each observer makes a given measurement choice. We also obtain  upper bounds on the required mutual information for arbitrarily biased measurement choices, that approach zero in the limit of extreme bias.

In Sec.~\ref{sec:singlet} we discuss the ordering of minimal mutual information with causal structure for maximally entangled two-qbuit states. Conclusions are presented in Sec.~\ref{sec:con}.

\section{Separable models of correlations}
\label{sec:sep}

Consider a given set of statistical correlations, represented by a set of joint probabilities $\{ p(a,b|x,y)\}$, where $(a,b)$ labels the outcomes of a joint experiment $(x,y)$ by two parties. We will use upper case letters to denote the corresponding random variables $A,B,X,Y$. Any underlying model of the correlations introduces a further random variable, $\Lambda$ (with specific values denoted by $\lambda$), on which the correlations depend. From Bayes theorem we have the identity
\beq \label{bayes}
p(a,b|x,y)  = \sum_\lambda  p(\lambda|x,y)\, p(a,b|x,y,\lambda),
\eeq
where summation is replaced by integration if one considers a continuous range of $\lambda$. Various classes of models are defined by imposing conditions on $p(\lambda|x,y)$ and $p(a,b|x,y,\lambda)$, and have been of deep interest in quantum foundations and information since Bell's seminal paper on local hidden variable models~\cite{bell1964}. 

This paper is concerned with models that satisfy the separability condition
\beq \label{sep}
p(a,b|x,y,\lambda) =p(a|x,\lambda)\, p(b|y,\lambda) .
\eeq 
Thus, conditioned on $\lambda$, the local measurement outcomes for each party are statistically independent of each other (`outcome independence'), and of the choice of measurement made by the other party (`parameter independence')~\cite{shimony84}. We use the term `separable' rather than `local' for models satisfying Eq.~(\ref{sep}), for reasons discussed at the end of this section.

It is well known that if, further, there is no correlation between the underlying variable $\lambda$ and the joint measurement $(x,y)$ (`measurement independence'), i.e., 
\beq \label{mind}
p(\lambda|x,y) = p(\lambda), 
\eeq
then the model is Bell local, and the correlations $\{ p(a,b|x,y)\}$  satisfy a corresponding set of Bell inequalities~\cite{bell1964,bellreview}.  The  predicted (and observed~\cite{aspect82,hanson15}) violation of Bell inequalities by some quantum systems implies that quantum mechanics is Bell nonlocal.

The assumption of measurement independence, as per Eq.~(\ref{mind}), is crucial to the security of device independent protocols in quantum information theory that rely on Bell nonlocality, as discussed in the Introduction. In particular, if measurement independence is relaxed, then there are underlying models which satisfy the separability condition~(\ref{sep}) but which violate Bell inequalities, in principle allowing  an adversary to determine the cryptographic key or random numbers generated by such protocols~\cite{kofler2006,koh2012}. 

The degree of measurement dependence in a given model is conveniently quantified by the mutual information shared between the measurement settings and the underlying variable~\cite{bg2011},
\beq \label{infgen}
I(X,Y:\Lambda) \coloneqq H(X,Y) - \sum_\lambda p(\lambda) H_\lambda(X,Y),
\eeq
where $H(X,Y) \coloneqq -\sum_{x,y} p(x,y) \log_2 p(x,y)$ denotes the entropy (in bits) of the distribution $p(x,y)$ of measurement settings (with summations replaced by integration over continuous ranges of $X,Y$ and/or $\Lambda$), and $H_\lambda(X,Y)$ denotes the entropy of the conditional distribution $p(x,y|\lambda)$. The mutual information measures the information cost of the correlation between the settings $(X,Y)$ and the underlying variable $\Lambda$, and vanishes only for models satisfying measurement independence as per Eq.~(\ref{mind}).

It has previously been shown that no more than \mbox{$\sim 0.066$} bits of measurement dependence is required for a separable model of a maximally entangled two-qubit state~\cite{relaxed}. Surprisingly, however, we will show that the corresponding model can in general only be implemented retrocausally, i.e., with future measurement settings affecting past source variables. Thus, particularly for practical implementations, it becomes necessary to distinguish different possible causal structures for measurement dependent models, and the information costs associated with them.

Two examples of causal structures for measurement dependent models are depicted in Fig.~\ref{fig:causvsretro}, with the random variables denoted by corresponding boxes, and causal influences by arrows. Figures~\ref{fig:causal} and~\ref{fig:retro} correspond to what we call `causal' and `retrocausal' measurement dependence, respectively. Such structures form directed acyclic graphs~\cite{pearl,wood2015,chaves2015}, and the examples in Fig.~\ref{fig:causvsretro} also appear in Figs.~26(b) and~27(b) of Ref.~\cite{wood2015}. There is a natural prescription for the form of the corresponding joint probability distribution of random variables $Z_1, Z_2,\dots,$ connected by a causal structure~\cite{pearl}:
\beq \label{prescription}
p(Z_1,Z_2,\dots) = \prod_j p(Z_j|{\rm Pa}(Z_j)),
\eeq
where ${\rm Pa}(Z_j)$ denotes the `parents' of $Z_j$, i.e, those random variables with arrows directly pointing to $Z_j$. This prescription is motivated by Reichenbach's principle that correlations arise only from direct or common causes~\cite{reich56,vanFraassen82}. For the examples in Fig.~\ref{fig:causvsretro} it implies, in particular, the separability condition~(\ref{sep}).  Note that it is also natural, in the implementation of quantum information protocols, for the measurement settings to appear to be selected randomly and independently. Hence, we will further typically impose the factorisability constraint
\beq \label{fact}
p(x,y) = p(x)\,p(y) 
\eeq
in what follows, unless explicitly indicated otherwise.

A main focus of this paper is to determine the minimum amount of measurement dependence needed for separable models of Bell nonlocality, under various causality constraints.
Fortunately, the task of finding an optimal separable model  is substantially simplified by noting we can, without any loss of generality, assume that the underlying outcome probabilities are deterministic, i.e., that
\beq \label{locdet}
p(a,b|x,y,\lambda)\in\{0,1\} .
\eeq
In particular, for any given nondeterministic model of a given set of joint correlations $\{p(a,b|x,y)\}$, one can construct a corresponding deterministic model which has the same degree of measurement dependence~\cite{relaxed}. This construction,  a generalisation of a one-party model by Bell~\cite{bell66}, has two additional variables, $\Lambda_A$ and $\Lambda_B$ say, held locally by each party so as to make their response functions deterministic. These additional variables are independent of the measurement $(x,y)$ and $\lambda$, implying that both models have the same value of the mutual information in Eq.~(\ref{infgen}), i.e., $I(X,Y:\Lambda,\Lambda_A,\Lambda_B)=I(X,Y:\Lambda)$~\cite{footdeterm}.  Thus, we can restrict attention to the class of deterministic separable models.

Finally, let us come back to our choice of terminology. We have preferred to use the neutral term `separable' for models satisfying condition~(\ref{sep}), rather than terminology involving `local'  (e.g., `measurement dependent locality'~\cite{putz2014}), for two main reasons. First, some models of this type can only be implemented retrocausally, as remarked above, which does not mesh with standard notions of locality. Second,  there are models of this type for {\it signaling} correlations~\cite{scarani,putz2014,hallbrans}, i.e., for correlations which do {\it not} satisfy
\beq \label{nonsig}
p(a|x,y) = p(a|x),~ p(b|x,y) = p(b|y) 
\eeq
for all $a,b,x$ and $y$, which leads to a conceptual tension for `locality' at the underlying and observable levels of such models (this is related to the subtlety of `signaling' for measurement dependent models; see discussions in Refs.~\cite{scarani,hallbrans,putz2016}).

\section{Optimal separable models for the CHSH scenario}
\label{sec:chsh}

In the CHSH scenario two observers, Alice and Bob say, can each make one of two binary measurements. For convenience we will label  their choice of measurement by $x,y = 0,1$, and their outcomes by $a,b=\pm1$. For any separable model, it follows from Eqs.~(\ref{bayes}) and~(\ref{sep}) that the average joint correlation for the pair of measurements $(x,y)$ is given by
\begin{align} \label{abav} 
\langle A B\rangle_{xy} & \coloneqq \sum_{a,b,\lambda} ab\, p(a,b,\lambda|x,y) 
 \nn \\
&= \sum_\lambda  p(\lambda|x,y) \, A_x(\lambda) B_y(\lambda),
\end{align}
where 
\beq \label{abdef}
A_x(\lambda)\coloneqq \sum_a a \,p(a|x,\lambda),~ B_y(\lambda)\coloneqq \sum_b b\,p(b|y,\lambda) 
\eeq
denote the average expectation values of $a$ and $b$ for a given value of $\lambda$ and  measurement settings $x,y$, and summation over $\lambda$ is replaced by integration over any continuous range of $\lambda$. 

If measurement independence is also satisfied, as per Eq.~(\ref{mind}),
then the correlations are Bell local and satisfy the well-known CHSH Bell inequalities, given by~\cite{chsh,bellreview}
\beq \label{chsh}
S\coloneqq  \langle AB\rangle_{00} +\langle AB\rangle_{01} + \langle AB\rangle_{10} - \langle AB\rangle_{11} \leq 2
\eeq
and the seven distinct permutations obtained therefrom by swapping the signs of the outcomes corresponding to one or more measurement settings.
However, if measurement independence is relaxed, then there are separable models for which the CHSH parameter $S$ in Eq.~(\ref{chsh}) can be as large as 4~\cite{hall2010, bg2011}, corresponding to the maximum possible algebraic value of $S$ (obtained for $ \langle AB\rangle_{xy} = (-1)^{xy}$~\cite{rastall85}). We will thus be interested in $S\in[2,4]$, with any value $S > 2$ corresponding to a violation of the CHSH inequality~\eqref{chsh}. Of particular interest is the value $S_Q \coloneqq 2\sqrt{2}$, which is the maximal possible value one can reach quantum mechanically (in a standard Bell test, with an independent choice of measurement settings)---the so-called Tsirelson bound~\cite{csirelson80}.

Our goal in this section is to find the minimum degree of measurement dependence required to violate the CHSH inequality~(\ref{chsh}) by a given amount, for various causal structures. As noted in the previous section, we may restrict attention to the class of deterministic models as per Eq.~(\ref{determ}), without any loss of generality, corresponding to deterministic outcomes,
\beq \label{determ}
A_x(\lambda),~ B_y(\lambda) = \pm 1 ,
\eeq
in Eq.~(\ref{abdef}).
Here $A_x(\lambda)$ and $B_y(\lambda)$ can be understood as the (deterministic) `response functions' of Alice and Bob.

It is natural to further restrict attention to models that generate nonsignaling correlations, as per Eq.~(\ref{nonsig}).
However, this is in fact a trivial constraint in the scenario we consider here: for any given deterministic separable model, one can always construct a corresponding deterministic separable model that is nonsignaling, and which has the same values of the CHSH parameter $S$ and mutual information $I$. This is achieved by taking an equal mixture of the given model with essentially the same model but with the outcomes flipped. More formally, introduce an additional (unbiased) source variable, $\Lambda'\in\{0,1\}$, and define $p(x,y,\lambda,\lambda')\coloneqq \half p(x,y,\lambda)$ and 
\beq \label{flip}
A_x(\lambda,\lambda')\coloneqq  (-1)^{\lambda'}A_x(\lambda),~ B_y(\lambda,\lambda')\coloneqq  (-1)^{\lambda'}B_y(\lambda).
\eeq  
This leaves the correlators $\langle A B\rangle_{xy}$ unchanged, which together with the independence of $\Lambda'$ from $X$ and $Y$ implies that $S$ and $I(X,Y:\Lambda,\Lambda') = I(X,Y:\Lambda)$ are invariant. Moreover, nonsignaling as per Eq.~(\ref{nonsig}) is satisfied, with $p(a|x)=p(b|y)=\half$. Hence, in our search for optimal models of Bell nonlocality having minimal measurement dependence we both can and will ignore the no-signaling constraint, since it can be trivially imposed via the above construction.

In the following subsections it is convenient to assume that the measurement settings appear (when not conditioned on $\lambda$) to be selected randomly and independently, i.e., 
\beq \label{uniform}
p(x,y) = \frac{1}{4},
\eeq
for all $x,y$ (so that $H(X,Y) = -\log_2 \frac14 = 2$), consistent with Eq.~(\ref{fact}). We will relax this condition in Sec.~\ref{sec:biased}.

\subsection{Optimal separable model with arbitrary measurement dependence}
\label{sub:gen}

We first consider the case where no constraints are placed on  measurement dependence, and determine the separable model that requires the least mutual information for a given CHSH violation. This model is related to the separable deterministic model of the singlet state given in Ref.~\cite{hall2010} (see Sec.~\ref{sec:singlet} below), and its optimality confirms a conjecture made in~\cite{fried2019}. It turns out, somewhat surprisingly, that this model can only be implemented via retrocausal measurement dependence, as per Fig.~\ref{fig:retro}.

From Bayes theorem we have that
\beq \label{bayes2}
p(\lambda|x,y) = \frac{p(\lambda) p(x,y|\lambda)}{p(x,y)}.
\eeq
Noting that $S$ in Eq.~(\ref{chsh}) can be written as
\beq
S = \sum_{x,y} (-1)^{xy}\langle AB\rangle_{xy},
\eeq
it then follows via Eqs.~(\ref{abav}) and~(\ref{uniform}) that
\begin{align}
S= & \ 4 \sum_{x,y,\lambda} (-1)^{xy} p(\lambda) \, p(x,y|\lambda) A_x(\lambda) B_y(\lambda).
\end{align}
It is convenient here to classify the possible values of $\lambda$, depending on whether $A_x(\lambda)$ takes the same value for $x=0,1$, and similarly for $B_y(\lambda)$. Let us thus define the sets 
\begin{align}
\mathfrak{L}_{\mu\nu} \coloneqq  \big\{ \lambda : \ & A_1(\lambda)=(-1)^\mu A_0(\lambda), 
 B_1(\lambda)=(-1)^\nu B_0(\lambda)\big\},
\end{align}
for $\mu,\nu=0,1$. Hence, for $\lambda \in \mathfrak{L}_{\mu\nu}$ and $x,y=0,1$,
\begin{align} \label{Ax_A0_By_B0}
A_x(\lambda)=(-1)^{\mu x} A_0(\lambda) \ \textrm{and} \ B_y(\lambda)=(-1)^{\nu y} B_0(\lambda).
\end{align}
The expression for $S$ above can then be written as
\begin{align}
S= 4 \sum_{\mu,\nu}\!\sum_{\lambda\in\mathfrak{L}_{\mu\nu}} \!\!\! p(\lambda) A_0(\lambda) B_0(\lambda) \sum_{x,y} (-1)^{xy + \mu x + \nu y} p(x,y|\lambda).
\end{align}
We may now substitute $\sum_{x,y} (-1)^{xy + \mu x + \nu y} p(x,y|\lambda) = (-1)^{\mu\nu} \sum_{x,y} (-1)^{(x+\nu)(y+\mu)} p(x,y|\lambda) = (-1)^{\mu\nu} [1-2p(x{=}\bar\nu,y{=}\bar\mu|\lambda)]$ (using the identity $\sum_{x,y} p(x,y|\lambda) = 1$), with $\bar\mu \coloneqq 1{-}\mu$ and $\bar\nu \coloneqq 1{-}\nu$, to give
\begin{align} \label{s_expl_calc}
& S= 4 \sum_{\mu,\nu}\sum_{\lambda\in\mathfrak{L}_{\mu\nu}} p(\lambda) A_0(\lambda) B_0(\lambda) \nn \\[-3mm]
& \hspace{30mm} (-1)^{\mu\nu} \big[1-2p(x{=}\bar\nu,y{=}\bar\mu|\lambda)\big].
\end{align}
It immediately follows that one has the tight bound
\beq \label{stight}
S\leq 4\sum_{\mu,\nu}\sum_{\lambda\in\mathfrak{L}_{\mu\nu}}  p(\lambda) \big|1-2p(x{=}\bar\nu,y{=}\bar\mu|\lambda) \big|
\eeq
for $S$, with saturation achieved by  choosing $A_0(\lambda)$ and $B_0(\lambda)$ such that 
\beq \label{sat1}
A_0(\lambda)B_0(\lambda) = (-1)^{\mu\nu}\,{\rm sgn}\big[1- 2p(x{=}\bar\nu,y{=}\bar\mu|\lambda)\big]
\eeq
for $\lambda\in\mathfrak{L}_{\mu\nu}$.
Inequality~(\ref{stight}) is central to obtaining the main results of this and the following subsections.

Further, if $p_{\min}\, (\le \frac14)$ denotes the infimum of $p(x,y|\lambda)$ over all values of $x,y$ and $\lambda$, then we have $\big|1-2p(x{=}\bar\nu,y{=}\bar\mu|\lambda) \big| \le 1-2p_{\min}$, which yields, via Eq.~(\ref{stight}), the tight bound
\beq \label{chshmod}
S\leq 4  \sum_\lambda   p(\lambda) (1-2p_{\min}) = 4 - 8p_{\min},
\eeq
with saturation achieved by choosing 
\beq \label{sat2}
p(x{=}\bar\nu,y{=}\bar\mu|\lambda) = p_{\min} ~{\rm or}~1-p_{\min}
\eeq
for $\lambda\in\mathfrak{L}_{\mu\nu}$ and the measurement outcomes as per Eq.~(\ref{sat1}).  Note that Eq.~(\ref{chshmod}) implies the relaxed CHSH inequalities in Eq.~(4) of Ref.~\cite{koh2012} (with $G=1$ and $P=\max_{x,y,\lambda} p(x,y|\lambda) \ge (1-p_{\min})/3$), and is equivalent to Eq.~(11) of Ref.~\cite{putz2014} (with $\ell'= p_{\min}$).

To find the minimal mutual information cost for a given value $S$ of the CHSH parameter, note first that Eq.~\eqref{chshmod} above implies that
\beq \label{pmin}
p_{\min} \leq (4-S)/8.
\eeq
Further, the mutual information simplifies via Eqs.~(\ref{infgen}) and~(\ref{uniform}) to
\beq \label{infret}
I(X,Y:\Lambda) = 2 - \sum_\lambda p(\lambda) H_\lambda(X,Y),
\eeq
and hence is minimised by maximising $H_\lambda(X,Y)$ for each $\lambda$, i.e., by making $p(x,y|\lambda)$ as uniform as possible. In turn, this is achieved by taking $p_{\min} (\in [0,\frac14])$ as large as possible, i.e, saturating Eq.~(\ref{pmin}), and $p(x,y|\lambda)$ to be a  distribution of the form 
\beq \label{retro_distr}
\{p_{\min}, (1-p_{\min})/3, (1-p_{\min})/3,(1-p_{\min})/3\} 
\eeq
up to permutations~\cite{footentmax}. Note this corresponds to taking the first choice in Eq.~(\ref{sat2}), since the second choice forces a distribution of the form $\{p_{\min}, 1-p_{\min},0,0\}$.
By considering all 4 permutations of the distribution in Eq.~\eqref{retro_distr}, corresponding to the first choice in Eq.~(\ref{sat2}) for the 4 respective cases $\lambda\in\mathfrak{L}_{\mu\nu}$ (with each giving the same value of $H_\lambda(X,Y)$ in Eq.~\eqref{infret}), one can further ensure that the average distribution of settings $p(x,y) = \sum_\lambda p(\lambda)p(x,y|\lambda) =\frac14$ as per Eq.~\eqref{uniform}: see the explicit model below.

\begin{figure}[!t] 
	\centering
	\includegraphics[width=0.45\textwidth]{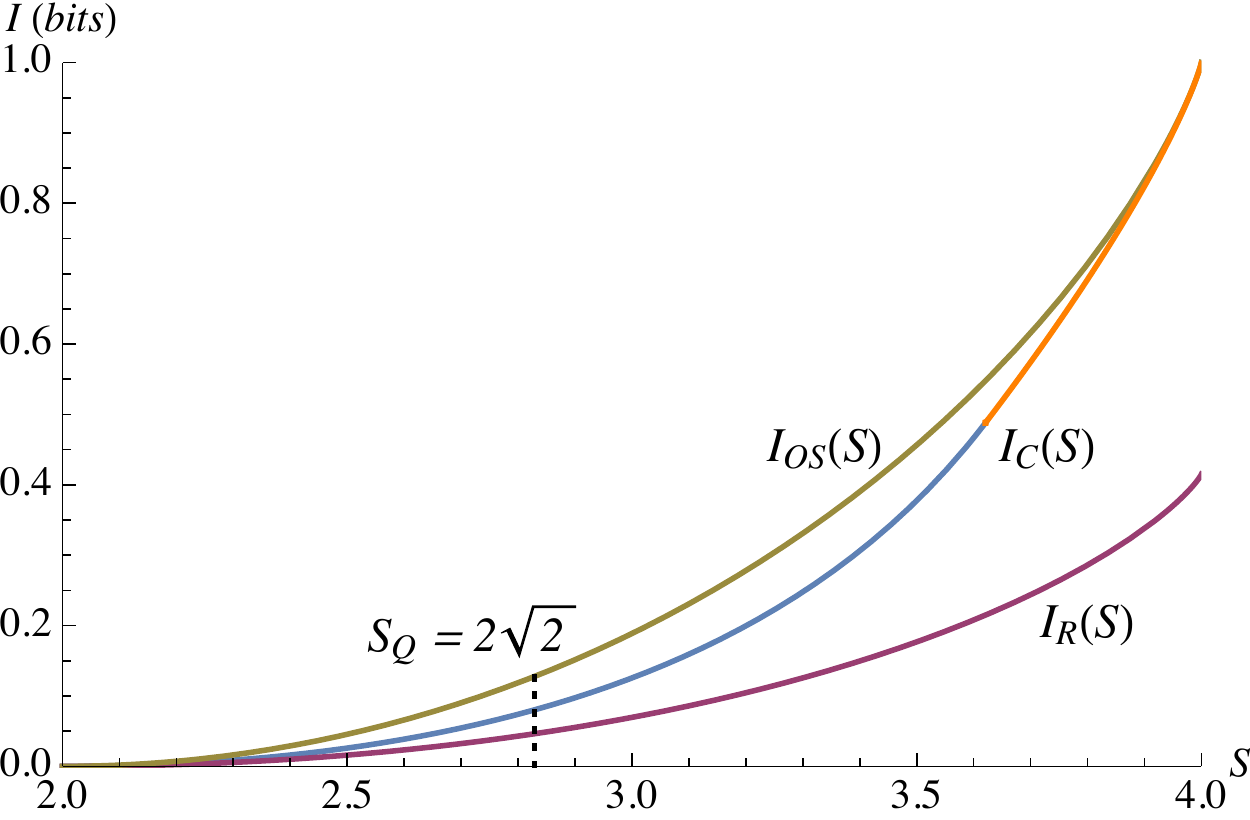}
	\caption{Plot of the minimal mutual information between source and measurement setting variables required to reach a value $S \in [2,4]$ of the CHSH parameter, for various causal structures. Retrocausal models are bounded below by the magenta curve [$I_R(S)$ in Eq.~\eqref{imin}], causal models by the blue and orange curve [$I_C(S)$ in Eq.~\eqref{icausal}; see also Fig.~\ref{fig:I1_I2} further below],  and one-sided models by the olive curve [$I_{OS}(S)$ in Eq.~\eqref{ios}]. Note that $I_R(S) \leq I_C(S)\leq I_{OS}(S)$, i.e., retrocausal models are more efficient than causal models, which in turn are more efficient than one-sided models. The value $S_Q=2\sqrt{2}$ indicates the Tsirelson bound, i.e., the maximal quantum violation of the CHSH inequality (with independent measurement settings).
	}
	\label{fig:I_vs_S}
\end{figure}

The corresponding minimum mutual information possible, for a given value $S$, follows from Eq.~(\ref{infret}) as
\begin{align} \label{imin}
I_R(S) \coloneqq  2 -h\left( \frac{4-S}{8} \right) - \frac{4+S}{8} \log_2 3 ,
\end{align}
where 
\begin{align} \label{hp}
h(p) \coloneqq  -p \log_2 p - (1-p) \log_2(1-p)
\end{align}
is the binary entropy function. Here the subscript $R$ stands for `retrocausal' because, as will be seen, this minimum value can only be obtained by retrocausal models. This is plotted in Fig.~\ref{fig:I_vs_S}, and ranges from 0 bits for $S = 2$ (no violation of the CHSH Bell inequality) to $\log_2\tfrac{4}{3}\sim 0.415$ bits for $S=4$ (maximum algebraic violation). For the maximum quantum violation, $S_Q=2\sqrt{2}$, one has
\beq \label{irsq}
I_R(S_Q) \sim 0.046~{\rm bits},
\eeq
i.e., just a little more than $1/22$ of a bit of measurement dependence is required to reach the Tsirelson bound.

The simplest optimal separable model reaching the minimal mutual information of Eq.~\eqref{imin}, satisfying the saturating conditions~(\ref{sat1}) and~(\ref{sat2}), with equality in Eqs.~(\ref{pmin}) and with uniform settings $x,y$ as in Eq.~\eqref{uniform}, is obtained by taking the four sets $\mathfrak{L}_{\mu\nu}$ to contain just one element $\lambda_{\mu\nu}$ each, with equal probabilities $p(\lambda_{\mu\nu})=1/4$. This model  is given explicitly in Table~\ref{tab1}. It is equivalent to the separable model in Tables~I and~II of Ref.~\cite{fried2019} (with $p_1=p_2=(1-4p)/3$ and $p_3=0$), which was conjectured to be optimal; our analysis above thus allows us to prove that conjecture. Note that this model gives signaling correlations; but as explained earlier with reference to Eq.~(\ref{flip}), it can easily be extended to a (still optimal) nonsignaling model that gives the same CHSH value $S$ for the same amount of measurement dependence.

\begin{table}[!t]
	
	\begin{ruledtabular}
		\begin{tabular}{c|c|cccc|cccc}
			$\lambda$ & $p(\lambda)$ & $p_{00|\lambda}$ & $p_{01|\lambda}$ & $p_{10|\lambda}$ & $p_{11|\lambda}$ & $A_0(\lambda)$ & $A_1(\lambda)$ & $B_0(\lambda)$ & $B_1(\lambda)$\\[1mm]
			\hline
			$\lambda_{00}$ & $1/4$ & $\frac{1{-}p}{3}$ & $\frac{1{-}p}{3}$ & $\frac{1{-}p}{3}$ & $p$ & $s$ & $s$ & $s$ & $s$ \\[1mm]
			$\lambda_{10}$ & $1/4$ & $\frac{1{-}p}{3}$ & $\frac{1{-}p}{3}$ & $p$ & $\frac{1{-}p}{3}$ & $t$ & $-t$ & $t$ & $t$ \\[1mm]
			$\lambda_{01}$ & $1/4$ & $\frac{1{-}p}{3}$ & $p$ & $\frac{1{-}p}{3}$ & $\frac{1{-}p}{3}$ & $u$ & $u$ & $u$ & $-u$ \\[1mm]
			$\lambda_{11}$ & $1/4$ & $p$ & $\frac{1{-}p}{3}$ & $\frac{1{-}p}{3}$ & $\frac{1{-}p}{3}$ & $v$ & $-v$ & $-v$ & $v$
		\end{tabular}
	\end{ruledtabular}
	\caption{\label{tab1} Optimal separable model, with the minimal required measurement dependence required to obtain a given value of the CHSH parameter $S$. The probabilities $p_{xy|\lambda}\coloneqq  p(x,y|\lambda)$ are defined for $\lambda=\lambda_{\mu\nu}$ such that $p_{xy|\lambda} = p  \in[0,1/4]$ if $x=\bar\nu$ and $y=\bar\mu$, and $p_{xy|\lambda} = \frac{1-p}{3} \in [\frac14,\frac13]$ otherwise. Thus, $p\equiv p_{\min}$ in Eq.~(\ref{retro_distr}). With $p(\lambda) \equiv \frac14$, the average probability distribution of the joint measurement setting $(x,y)$ is $p(x,y) = \sum_\lambda p(\lambda)p_{xy|\lambda} = \frac14$, as required by Eq.~\eqref{uniform}, and the `retrocausal' conditional probabilities $p(\lambda|x,y)$ are $p(\lambda|x,y) = p(x,y|\lambda)p(\lambda)/p(x,y) = p_{xy|\lambda}$. The parameters $s,t,u,v$ defining the measurement outcomes of the model can take any value $\pm 1$. These response functions are defined, in accordance with Eqs.~\eqref{Ax_A0_By_B0} and~\eqref{sat1} (noticing that all $p_{xy|\lambda} \le \half$), such that $A_x(\lambda_{\mu\nu})B_y(\lambda_{\mu\nu}) = (-1)^{\mu x+\nu y+\mu\nu}$. One thus obtains $\langle A B\rangle_{xy} = \sum_\lambda p(\lambda|x,y) A_x(\lambda)B_y(\lambda) = (-1)^{xy}(1-2p)$. The corresponding value of the CHSH parameter is $S=4-8p$, with corresponding mutual information given by Eq.~(\ref{imin}). As shown in the main text, this model has inherent retrocausal measurement dependence.}
\end{table}

There is clearly a retrocausal implementation of the model in Table~\ref{tab1} (and indeed of any measurement dependent model), as per Fig.~\ref{fig:retro}, where the source receives the values $x$ and $y$ of $X$ and $Y$ with some prior factorised probability $p(x,y)=p(x)p(y)$ as per Eq.~(\ref{fact}), and generates $\Lambda=\lambda$ with probability $p(\lambda|x,y) =p(x,y|\lambda)p(\lambda)/p(x,y)$~\cite{footretro}.  However, it is of interest to ask whether the model in Table~\ref{tab1} also has a causal implementation, with the future measurement settings influenced by past source variables? As shown in the next section, the answer is negative: the model in Table~\ref{tab1} is {\it inherently retrocausal}.

\subsection{Optimal separable model with causal measurement dependence}
\label{sec:causal}

The structure of causal measurement dependent models is shown in Fig.~\ref{fig:causal}, where the value of the source variable $\lambda$ can causally influence the subsequent selection of measurements by Alice and Bob. Thus, all arrows in Fig.~\ref{fig:causal} are implemented in timelike directions.
Such models are characterised by the condition
\beq \label{causalmd}
p(x,y|\lambda) = p(x|\lambda) p(y|\lambda),
\eeq
following from the prescription in Eq.~(\ref{prescription}). This is analogous to the separability condition in Eq.~(\ref{sep}) (see also Ref.~\cite{koh2012}).
Note that no further generality is gained by writing
\beq \label{causal_extended}
p(x,y|\lambda) = \sum_{\lambda'} p(\lambda'|\lambda) p(x|\lambda,\lambda') p(y|\lambda,\lambda'),
\eeq
via an additional source variable $\Lambda'$ (which is always formally possible), as this merely generates a correlation model of the same form as Eqs.~(\ref{abav}) and~(\ref{causalmd}) with respect to the extended source variable $\Lambda''\coloneqq (\Lambda,\Lambda')$.

We call the condition in Eq.~(\ref{causalmd}) `causal measurement dependence' (it has also been called `independent sources'~\cite{putz2016}). It is a nontrivial constraint, leading to the requirement of a higher degree of mutual information to achieve a given Bell inequality violation than is the case for general separable models. In particular, it will be seen that the optimal model in Table~\ref{tab1} cannot be implemented via causal measurement dependence, but requires {\it retrocausal} measurement dependence as per Fig.~\ref{fig:retro} (or, among the other causal structures we shall consider, the supplemented `zigzag' measurement dependence of Fig.~\ref{fig:zigzag2} below, which also involves a retrocausal influence).

To determine the minimum mutual information $I(X,Y:\Lambda)$ required to simulate a violation of the CHSH inequality, under causal measurement dependence, necessitates a little more work than for the general case in Sec.~\ref{sub:gen}. First, following the previous analysis, one obtains a tight bound as per Eq.~(\ref{stight}) for the CHSH parameter as before, which is still saturated by deterministic measurement outcomes satisfying Eq.~(\ref{sat1}).
Second, denoting by $p_{\min}^X(\lambda)$ the minimal value of $p(x|\lambda)$ for $x=0,1$ and by $p_{\min}^Y(\lambda)$ the minimal value of $p(y|\lambda)$ for $y=0,1$ (both for a given $\lambda$), we have, under the assumption that $p(x,y|\lambda)$ decomposes as in Eq.~\eqref{causalmd}, that $p(x{=}\bar\nu,y{=}\bar\mu|\lambda), 1- p(x{=}\bar\nu,y{=}\bar\mu|\lambda)\ge p_{\min}^X(\lambda) \, p_{\min}^Y(\lambda)$, so that $\big|1-2p(x{=}\bar\nu,y{=}\bar\mu|\lambda) \big| \le 1 - 2 p_{\min}^X(\lambda) \, p_{\min}^Y(\lambda)$. Substituting this into Eq.~(\ref{stight}) then leads to the tight bound
\begin{align} \label{bnd_S_causal}
& S \leq \overline{S}_{\max} \coloneqq  \sum_\lambda p(\lambda) S_{\max}(\lambda) \nn \\
& \quad \textrm{with} \quad S_{\max}(\lambda) \coloneqq  4 - 8 \,  p_{\min}^X(\lambda) \, p_{\min}^Y(\lambda),
\end{align}
for causal models (rather than Eq.~\eqref{chshmod} for general models), with saturation achieved when
\beq
p(x{=}\bar\nu|\lambda) = p_{\min}^X(\lambda),~~p(y{=}\bar\mu|\lambda) = p_{\min}^Y(\lambda) \label{eq:saturate_causal}
\eeq
for $\lambda\in\mathfrak{L}_{\mu\nu}$.

The mutual information then follows via Eqs.~(\ref{infgen}), (\ref{uniform}) and~(\ref{causalmd}) as
\begin{align} \label{icausal_1}
I(X,Y:\Lambda) &= 2 -  \sum_{\lambda} p(\lambda) \big[ H_\lambda(X) + H_\lambda(Y) \big]\nn\\
&= \sum_\lambda p(\lambda) I(\lambda),
\end{align}
where
\begin{align} \label{ilam}
I(\lambda)&\coloneqq  2 -  h\left(p_{\min}^X(\lambda)\right) - h\left(p_{\min}^Y(\lambda)\right) .
\end{align}

Note that by letting one party introduce some local noise, one can obtain any value of $S$ between $0$ and $\overline{S}_{\max}$ above, without changing the mutual information.
This implies, in particular, that the minimal mutual information required to get a given value $S$ increases with $S$, and that it is indeed obtained by saturating the upper bound in Eq.~\eqref{bnd_S_causal}---i.e., by taking $S = \overline{S}_{\max} = \sum_\lambda p(\lambda) S_{\max}(\lambda)$.
This, together with Eq.~\eqref{icausal_1}, then implies that the optimal pairs of values $(S,I)$ are obtained as convex combinations of optimal pairs $(S_{\max}(\lambda), I(\lambda))$, for some fixed $\lambda$ (and with the weights $p(\lambda)$ in the combination taken so as to satisfy $p(x,y) = \sum_\lambda p(\lambda) p(x,y|\lambda) = \frac14$ as in Eq.~\eqref{uniform}).

\begin{figure}[!t] 
	\centering
	\includegraphics[width=0.45\textwidth]{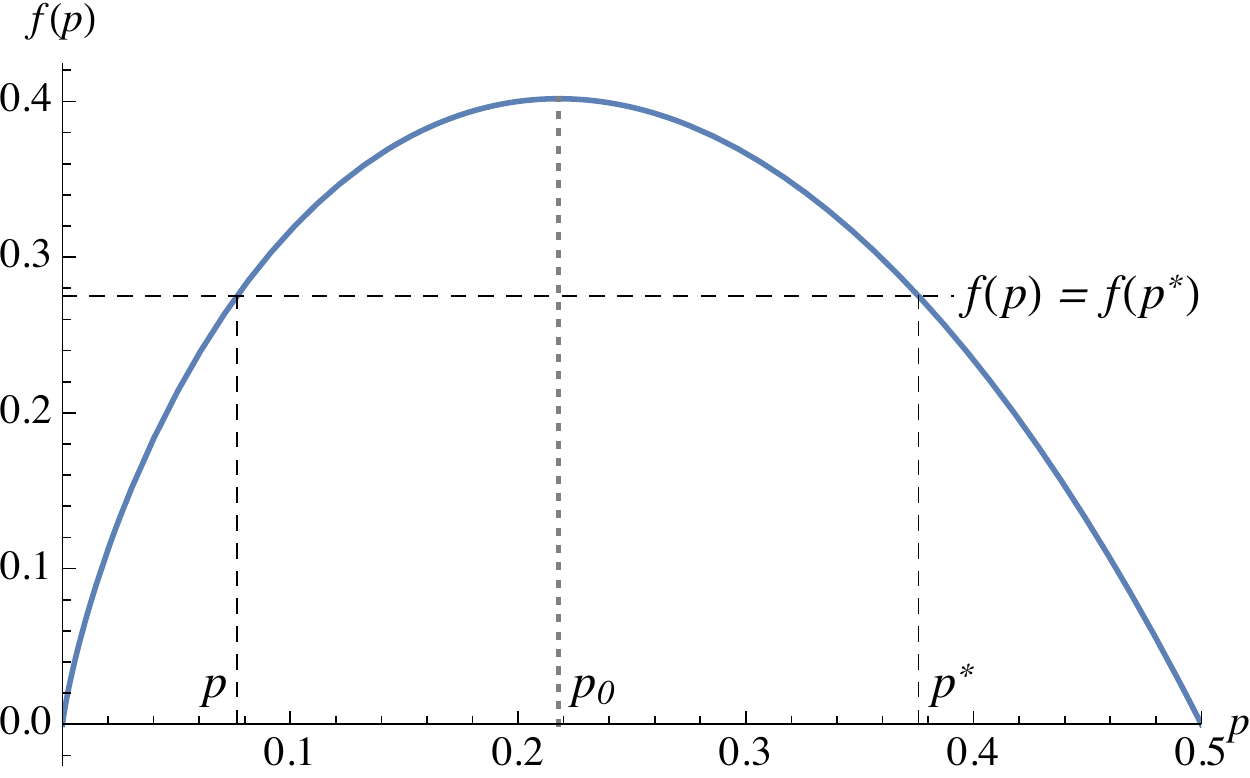}
	\caption{Function $f(p) \coloneqq p \log_2\frac{1-p}{p}$ plotted for $p\in[0,\half]$. There are in general two different values $p, p^*$ giving the same $f(p)=f(p^*)$. 
	}
	\label{fig:function}
\end{figure}

Note that both $S_{\max}(\lambda)$ and $I(\lambda)$ are expressed above as functions of $p_{\min}^X(\lambda)$ and $p_{\min}^Y(\lambda)$, which we shall simply denote here by $p_X$ and $p_Y$, resp., so as to lighten the notations: $S_{\max}(\lambda) = S_{\max}(p_X,p_Y) = 4 - 8p_Xp_Y$ and $I(\lambda) = I(p_X,p_Y) = 2-h(p_X)-h(p_Y)$. To find the optimal pairs $(S_{\max}(\lambda), I(\lambda))$, our goal is thus to calculate
\begin{align}
I_C(S) \coloneqq \ & \min_{p_X,p_Y \in [0,\half]} I(p_X,p_Y) \nn \\
& \text{s.t. } S_{\max}(p_X,p_Y) = S.
\end{align}
For this we introduce a Lagrange multiplier $\kappa$ and define the Lagrangian
\begin{align}
{\cal L}(p_X,p_Y,\kappa) \coloneqq I(p_X,p_Y) - \kappa \, [S_{\max}(p_X,p_Y) -S].
\end{align}
Setting $\partial {\cal L}/\partial p_X=\partial {\cal L}/\partial p_Y=0$ and eliminating $\kappa$ gives $f(p_X)=f(p_Y)$, where the function
\beq \label{func}
f(p) \coloneqq  p \log_2 {\textstyle \frac{1-p}{p}}
\eeq
is plotted in Fig.~\ref{fig:function}, for $p \in [0,\half]$.

As one can see, for any given value of $p_X$ there are generally two solutions to $f(p_X)=f(p_Y)$: (i) $p_Y=p_X$, and (ii) $p_Y=p_X^*$, where $p_X^*$ denotes the abscissa of the second point of intersection of the curve $f(p)$ with the horizontal line passing through $(p_X,f(p_X))$.
The  first solution gives, via Eq.~\eqref{bnd_S_causal}, $S_{\max}(\lambda) = 4 - 8 \, (p_X)^2 \in [2,4]$, and yields via Eq.~(\ref{ilam}) the associated mutual information $I(\lambda)=I_1(S_{\max}(\lambda))$, with
\beq \label{i1s}
 I_1(S) \coloneqq  2 - 2h\left(\sqrt{\frac{4-S}{8}}\right) .
\eeq
For the second solution, one has $S_{\max}(\lambda) = 4 - 8 \, p_X \, p_X^*$ from Eq.~(\ref{bnd_S_causal}). We find that this can only give values $S_{\max}(\lambda) \ge S_0 \sim 3.620$, where $S_0 \coloneqq 4 - 8 p_0^2$ is obtained for the value $p_0 \sim 0.218$ that gives the maximum of the function $f(p)$ (i.e., the solution of $\frac{\textup{d}f}{\textup{d}p} = \log_2 \big(\frac{1-p}{p}\big) - \frac{1}{(1-p)\log_e 2} = 0$, such that $p_0^*=p_0$; see Fig.~\ref{fig:function}). For a value of $S \ge S_0$, the solutions $p_X(S)$ and $p_X^*(S)$ to $S = 4 - 8 p_X \, p_X^*$ must in general (except for the extremal values $S = S_0$ and $S = 4$) be found numerically. The corresponding mutual information is then, according to Eq.~(\ref{ilam}), $I(\lambda)=I_2(S_{\max}(\lambda))$, where
\beq \label{i2s}
I_2(S) \coloneqq  2 - h\big(p_X(S)\big) - h\big(p_X^*(S)\big).
\eeq

\begin{figure}[!t] 
	\centering
	\includegraphics[width=0.45\textwidth]{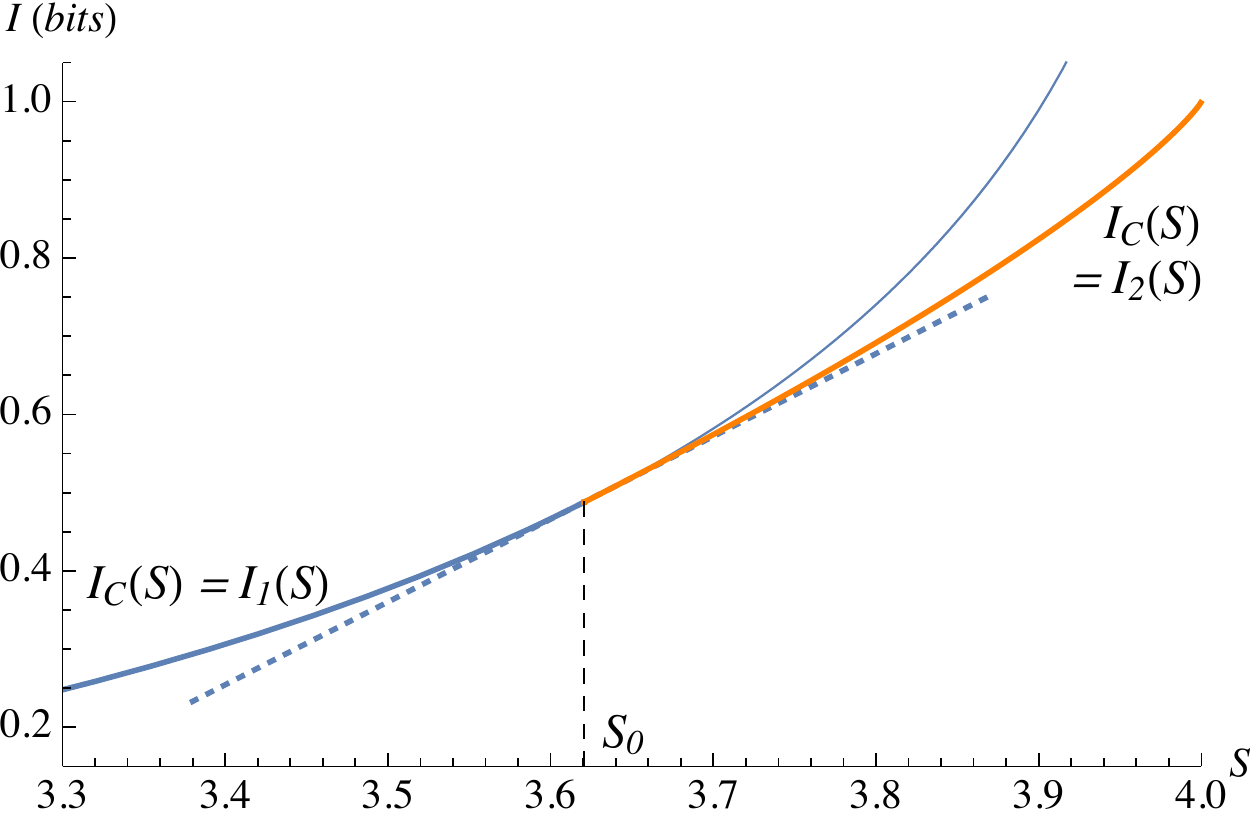}
	\caption{Plot of the functions $I_1(S)$ [blue curve, Eq.~\eqref{i1s}], $I_2(S)$ [orange curve, Eq.~\eqref{i2s}, defined for $S\ge S_0\sim 3.620$], and $I_C(S)$ [Eq.~\eqref{icausal}], for $S$ in the region around $S_0$. The dotted line represents the common tangent to $I_1(S)$ and $I_2(S)$ at $S=S_0$. A full plot of $I_C(S)$ is given in Fig.~\ref{fig:I_vs_S}.
	}
	\label{fig:I1_I2}
\end{figure}

The curves $I_1(S)$ (defined for any $S \in [2,4]$) and $I_2(S)$ (defined only for $S \ge S_0$) are plotted in Fig.~\ref{fig:I1_I2}, for the relevant values of $S$. One can see that $I_2(S) \le I_1(S)$ for $S \ge S_0$, so that the second solution to our optimisation problem provides a lower mutual information in that range. Furthermore, both $I_1(S)$ and $I_2(S)$ are seen to be convex, and one can prove that they share the same tangent at $S = S_0$ (see Appendix~\ref{app}), so that $\min[I_1(S),I_2(S)]$ is also convex. Taking convex combinations of $S_{\max}(\lambda)$ and $I(\lambda)$ as in Eqs.~\eqref{bnd_S_causal} and~\eqref{icausal_1} hence does not provide a lower mutual information. We will see with the explicit model below, however, that one can still combine different $\lambda$s having the same values of $S_{\max}(\lambda)$ and $I(\lambda)$, so as to satisfy $p(x,y) = \sum_\lambda p(\lambda)p(x,y|\lambda) =\frac14$ as required.

We conclude from this analysis that the minimum mutual information possible for a given value $S$ of the CHSH parameter, via a causally measurement dependent separable model, is given by
\begin{align} \label{icausal}
I_C(S) & = \left\{ \begin{array}{ll} 
I_1(S)  & \textrm{for} \ 2\leq S\leq S_0,\\
I_2(S) & \textrm{for} \ S_0 \leq S \le 4 .
\end{array} \right.
\end{align}
This is plotted in Fig.~\ref{fig:I_vs_S}, together with $I_R(S)$ from Eq.~(\ref{imin}). One finds that
\beq \label{icgtrir}
I_C(S) > I_R(S)
\eeq
for all $S>2$. Thus, as claimed earlier, the optimal separable model in Table~\ref{tab1} (whose input distribution $p(x,y|\lambda)$ indeed does not satisfy the causality condition of Eq.~\eqref{causalmd}) cannot be implemented via causal measurement dependence.

\begin{table}[!t]
	\begin{ruledtabular}
		\begin{tabular}{c|c|cc|cccc}
			$\lambda$ & $p(\lambda)$ & $p(x{=}0|\lambda)$ & $p(y{=}0|\lambda)$ & $A_0(\lambda)$ & $A_1(\lambda)$ & $B_0(\lambda)$ & $B_1(\lambda)$\\[1mm]
			\hline
			$\lambda_{00}$ & $1/4$ & $1{-}p$ & $1{-}\tilde p$ & $s$ & $s$ & $s$ & $s$ \\[1mm]
			$\lambda_{10}$ & $1/4$ & $1{-}p$ & $\tilde p$ & $t$ & $-t$ & $t$ & $t$ \\[1mm]
			$\lambda_{01}$ & $1/4$ & $p$ & $1{-}\tilde p$ & $u$ & $u$ & $u$ & $-u$ \\[1mm]
			$\lambda_{11}$ & $1/4$ & $p$ & $\tilde p$ & $v$ & $-v$ & $-v$ & $v$
		\end{tabular}
	\end{ruledtabular}
	\caption{\label{tab2} Optimal separable model with minimal causal measurement dependence, for any value $S \in [2,4]$ of the CHSH parameter. 
	The conditional probabilities $p(x|\lambda)$ and $p(y|\lambda)$ are defined, in terms of the parameter $p \in [0,\half]$, such that for $\lambda=\lambda_{\mu\nu}$, $p(x{=}0|\lambda)=1{-}p$ if $\nu=0$, $p(x{=}0|\lambda)=p$ otherwise, and $p(y{=}0|\lambda)=1{-}\tilde p$ if $\mu=0$, $p(y{=}0|\lambda)=\tilde p$ otherwise (as in Eq.~\eqref{eq:saturate_causal}). Here $\tilde p \coloneqq p$ if one aims at obtaining $S \le S_0$ (which includes any quantum violation), while $\tilde p \coloneqq p^*$ for $S \ge S_0$ (as defined in the main text). The values of $p(x,y|\lambda)$ may be determined from $p(x{=}0|\lambda)$ and $p(y{=}0|\lambda)$ via Eq.~(\ref{causalmd}) (and with $p(x(y){=}1|\lambda) = 1-p(x(y){=}0|\lambda)$). The average probability distribution of $x,y$ is $p(x,y) = \sum_\lambda p(\lambda)p(x,y|\lambda) = \frac14$, as required. The outcome parameters $s,t,u,v=\pm1$ are taken as in Table~\ref{tab1}, such that $A_x(\lambda_{\mu\nu})B_y(\lambda_{\mu\nu}) = (-1)^{\mu x+\nu y+\mu\nu}$. Noting that $p(\lambda|x,y)=p(x,y|\lambda)p(\lambda)/p(x,y)=p(x|\lambda)p(y|\lambda)$ here, one obtains $\langle A B\rangle_{xy} = (-1)^{xy}(1-2p\tilde p)$, and in turn $S = 4 - 8 p\tilde p$, with corresponding mutual information given by Eq.~(\ref{icausal}).}
\end{table}

An explicit optimal model that reaches the lower bound $I_C(S)$ above is given in Table~\ref{tab2}. Again, this model is signaling, but can easily be turned into a nonsignaling causal model which gives the same value of the CHSH parameter $S$, as per the construction in Eq.~(\ref{flip}), with the same amount of measurement dependence.

Note that the maximum quantum violation, $S_Q=2\sqrt{2}$, is strictly less than $S_0$. Hence, for any quantum violation the minimum mutual information can be calculated analytically from Eq.~(\ref{i1s}). One finds in particular that a minimum value
\beq \label{icsq}
I_C(S_Q)  =I_1(S_Q)\sim 0.080~{\rm bits}
\eeq
is required for any fully causal model reproducing the maximal quantum violation. Note this is nearly twice as much as $I_R(S_Q)$ in Eq.~(\ref{irsq}). Further, a fully causal model for the maximal algebraic value $S=4$ requires a mutual information of $I_C(4) = I_2(4) = 1$ bit (obtained for $p_X=0, p_X^* = \half$ in Eq.~(\ref{i2s})), i.e.,  more than twice as much as the value $I_R(4)=\log_2\tfrac{4}{3}\sim 0.415$ bits required for a retrocausal model.

\subsection{Optimal separable model with one-sided measurement dependence}
\label{sec:one-sided}

\begin{figure}[!t] 
	\centering
	\includegraphics[width=0.5\textwidth]{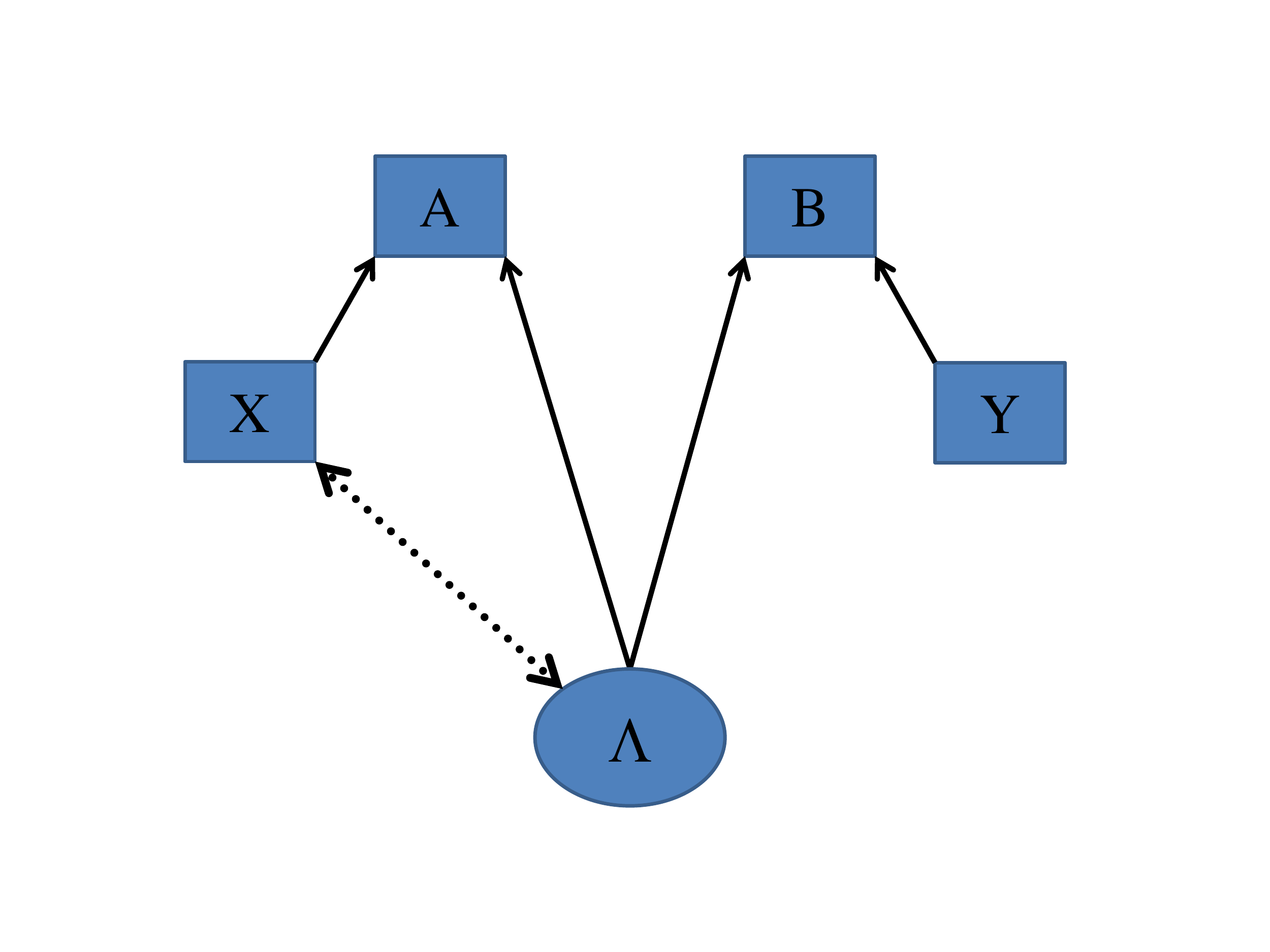}
	\caption{Two-party Bell scenario with  one-sided measurement dependence (dotted line), which may be either causal or retrocausal.
	}
	\label{fig:one-sided}
\end{figure}

We now consider the case of one-sided measurement dependence, where the source variable is correlated with just one of the measurement selections, as depicted in Fig.~\ref{fig:one-sided}. Taking this correlation to be on Alice's side, it follows that $Y$ is independent of $X$ and $\Lambda$, yielding 
\beq \label{onesided}
p(x,y|\lambda) = p(x|\lambda) \, p(y)
\eeq
for one-sided models. Thus, one-sided models are formally a special case of causal models as per Eq.~(\ref{causalmd}), implying they will, in general, require a greater degree of measurement dependence to achieve a given CHSH violation.
Note that one-sided models can equivalently be implemented either causally, via the earlier source variable $\lambda$ influencing the later measurement selection $x$, or retrocausally, with $x$ influencing $\lambda$, as indicated by the double arrow head in Fig.~\ref{fig:one-sided}. 

For the CHSH scenario, assuming unbiased measurement selection probabilities as per Eq.~(\ref{uniform}), summation over $x$ in Eq.~(\ref{onesided}) gives
\beq
p(y|\lambda) = p(y) = \half .
\eeq
 Hence, we have $p_{\min}^Y(\lambda) = \min_y p(y|\lambda) = \half$, and can directly use Eq.~(\ref{bnd_S_causal}) for causal models to write the maximum possible value of the CHSH parameter as
\begin{align}
S \leq \sum_\lambda p(\lambda) \big[ 4 - 4 \,  p_{\min}^X(\lambda) \big]
\end{align}
(with saturation achieved when $p(x{=}\bar\nu|\lambda) = p_{\min}^X(\lambda)$ for $\lambda\in\mathfrak{L}_{\mu\nu}$), so that
\begin{align}\label{bnd_avg_pmin_OS}
\sum_\lambda p(\lambda)p_{\min}^X(\lambda) \leq 1 - S/4.
\end{align}
The mutual information on the other hand is given via Eq.~(\ref{onesided}) by
\begin{align}
& \hspace{-2mm} I(X,Y:\Lambda) = I(X:\Lambda) = 1 - \sum_\lambda p(\lambda) \, h\left(p_{\min}^X(\lambda)\right) \nn\\
& \hspace{3mm} \geq 1 - h\left(\sum_\lambda p(\lambda) \, p_{\min}^X(\lambda)\right) \geq 1 - h\left(1-S/4\right) ,
\end{align}
where the first inequality follows from the concavity of the binary entropy function $h$ and is saturated when $p_{\min}^X(\lambda)$ is independent from $\lambda$, while the second one follows from Eq.~\eqref{bnd_avg_pmin_OS} above, and the fact that $h(p)$ increases monotonically for $p \in [0,\half]$. We thus find that the minimal required mutual information for one-sided measurement dependence is
\beq \label{ios}
I_{OS}(S) \coloneqq 1-h\left(\frac{S}{4}\right).
\eeq
A corresponding optimal model (which further satisfies $p(x) = \sum_\lambda p(\lambda) p(x|\lambda)=\half$, as required to recover Eq.~\eqref{uniform}), for all values of $S\in[2,4]$, is obtained by replacing $\tilde p$ with $\half$ in Table~\ref{tab2}. 

The mutual information in Eq.~(\ref{ios}) is plotted in Fig.~\ref{fig:I_vs_S}, and is seen to be larger than both $I_R(S)$ and $I_C(S)$, as expected.  For the maximum quantum violation $S_Q = 2\sqrt{2}$ one finds in particular that
\beq \label{iosvq}
I_{OS}(S_Q) \sim 0.128~ {\rm bits}.
\eeq
Note that this is approximately half of the mutual information of $\sim 0.247$~bits required in the Banik~{\it et~al.} one-sided model~\cite{banik12,fried2019} (although the latter model is optimal for the measure of measurement independence $M$ introduced in Ref.~\cite{hall2010}).
For the algebraic maximum $S=4$, one obtains $I_{OS}(4) = 1$ bit, which is the same value as $I_C(4) = 1$ bit for optimal causal models.

\subsection{Zigzag measurement dependence}
\label{sec:zigzag}

\begin{figure}[!t] 
	\centering
	\includegraphics[width=0.5\textwidth]{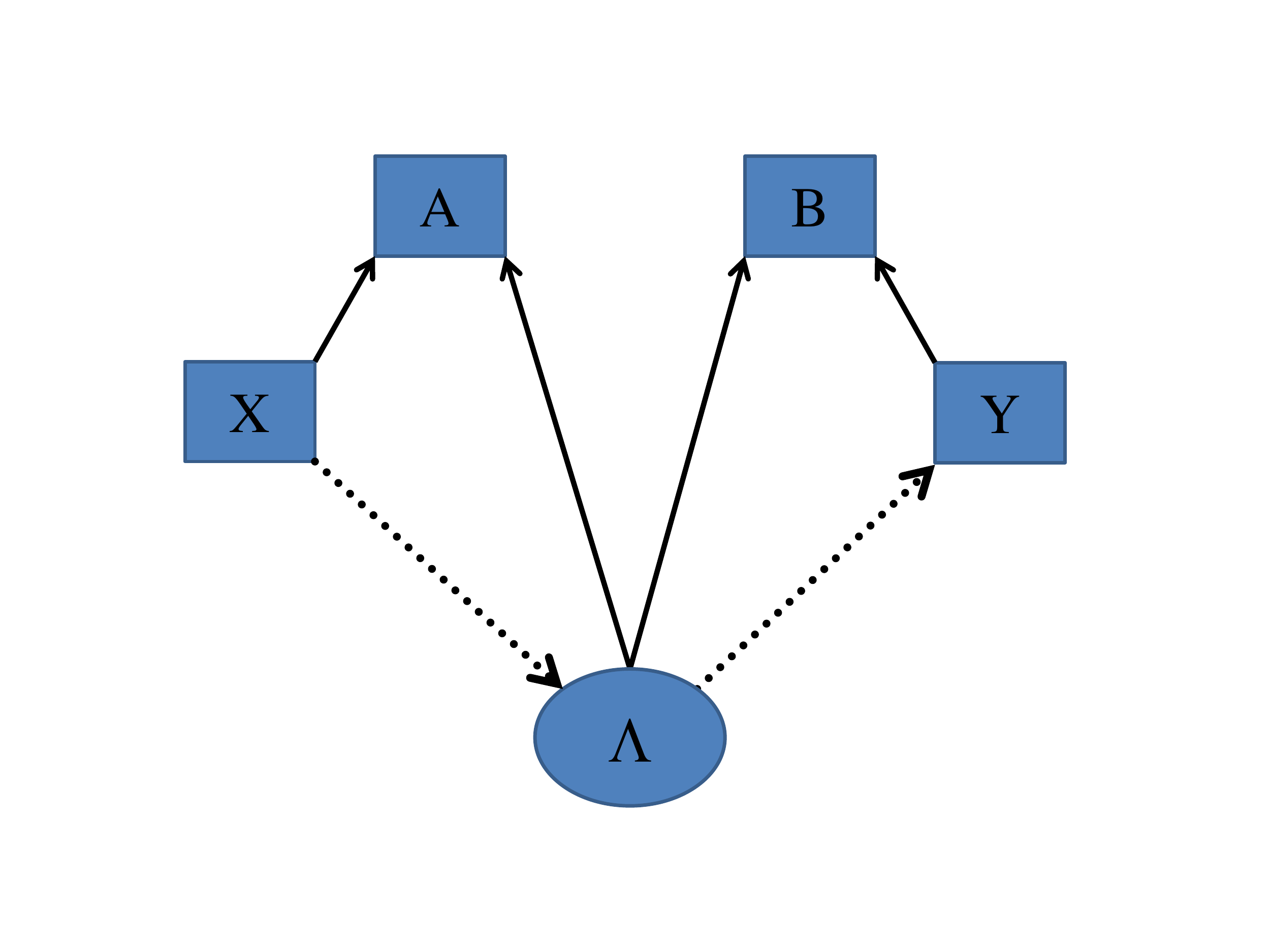}
	\caption{Two-party Bell scenario with zigzag measurement dependence (dotted arrows). As shown in the main text, this is formally equivalent to causal measurement dependence as per Fig.~\ref{fig:causal} (and it reduces to one-sided measurement dependence, as per Fig.~\ref{fig:one-sided}, if only one dotted arrow is permitted).
	}
	\label{fig:zigzag}
\end{figure}

A further type of causal structure that has been considered for the explanation of quantum correlations is zigzag causality, as depicted in Fig.~\ref{fig:zigzag}. This type of causal structure was introduced by Costa de~Beauregard~\cite{costa1953,costa1977}, and has recently been examined by Price and Wharton~\cite{price2015}. As shown in Fig.~\ref{fig:zigzag}, one of the measurement selections, Alice's say, influences the source variable,  which can in turn influence the other measurement selection.  

Applying the prescription in Eq.~(\ref{prescription}) to zigzag measurement dependence yields $p(x,y,\lambda) = p(x)p(y|\lambda)p(\lambda|x) =  p(x,\lambda) p(y|\lambda)$, reflecting the lack of a direct causal influence from $X$ to $Y$ in Fig.~\ref{fig:zigzag}. Thus, dividing by $p(\lambda)$,
\beq \label{zigzagmd}
p(x,y|\lambda) = p(x|\lambda)\, p(y|\lambda).
\eeq
Comparing with Eq.~(\ref{causalmd}) (and since these are, in both cases, the only further constraints imposed on separable models), it follows that zigzag measurement dependence is formally equivalent to causal measurement dependence, and so may be analysed precisely as in Sec.~\ref{sec:causal}. In particular, for the CHSH scenario with $p(x,y)=\frac14$, the corresponding minimal information is given by
\beq
I_Z(S) \coloneqq I_C(S),
\eeq
for zigzag models, where $I_C(S)$ is defined in Eq.~(\ref{icausal}). We note this equivalence can be seen as strengthening de~Beauregard's analogy between zigzag causality and particle-antiparticle pair creation~\cite{costa1953,costa1977}, for which causal propagation of the pair is formally equivalent to a single particle first propagating retrocausally then causally.

	\begin{figure}[!t] 
	\centering
	\includegraphics[width=0.5\textwidth]{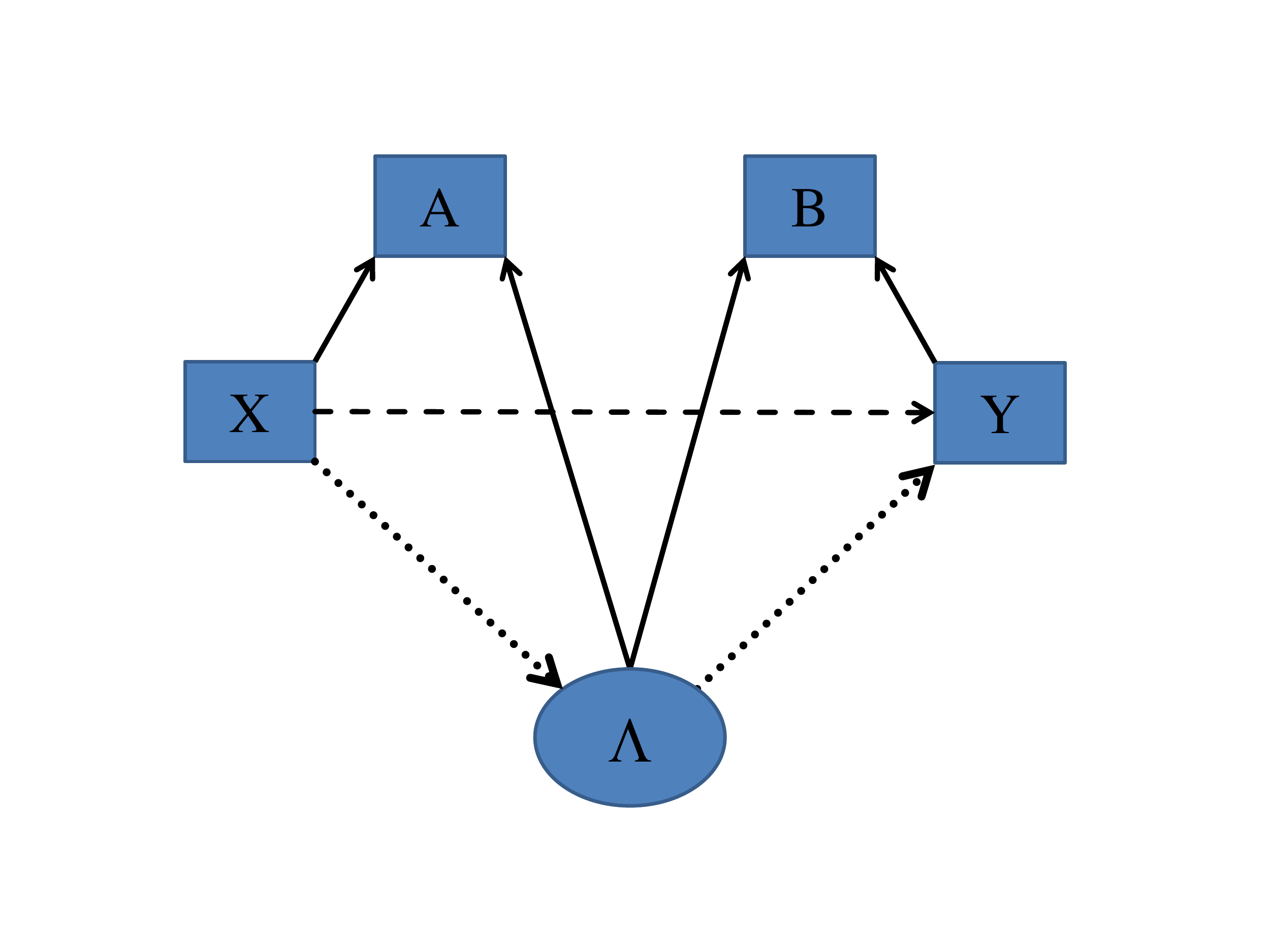}
	\caption{Zigzag measurement dependence supplemented with a superluminal influence from $X$ to $Y$ (dashed line). As shown in the main text, this scenario is formally equivalent to retrocausal measurement dependence as per Fig.~\ref{fig:retro}.
	}
	\label{fig:zigzag2}
\end{figure}

It is also of interest to note that if one supplements zigzag measurement dependence with a superluminal influence from $X$ to $Y$, as depicted in Fig.~\ref{fig:zigzag2}, then the prescription in Eq.~(\ref{prescription}) places no restrictions on the form of $p(x,y,\lambda)$. Hence, this modification is formally equivalent to the general case discussed in Sec.~\ref{sub:gen}, which we have in turn seen is formally equivalent to retrocausal measurement dependence as per Fig.~\ref{fig:retro}. For the CHSH scenario this implies that adding a superluminal arrow to zigzag measurement dependence as per Fig.~\ref{fig:zigzag2} decreases the minimal mutual information required, from $I_Z(S)$ to $I_R(S)$, where the latter is defined in Eq.~(\ref{imin}).

\subsection{Superdeterministic models}

Superdeterminism is a rather strong form of measurement dependence, in which both the measurement settings and measurement outcomes are fully determined by underlying variables~\cite{superdeterm,superdeterm1,hooft2014,bigbelltest,superdeterm2}. The latter variables themselves must still retain a statistical quality (corresponding to ignorance of initial conditions), for superdeterministic models to be able to reproduce quantum predictions. The causal structure of a superdeterministic model corresponds, therefore, to causal measurement dependence as in Fig.~\ref{fig:causal}, but with the values of each of $A,B, X$ and $Y$ being fully determined by the value of $\Lambda$.

For the CHSH scenario it follows that $p(x,y|\lambda)\in\{0,1\}$ for  superdeterministic models. Hence,  the mutual information of any such model follows via Eq.~(\ref{infgen}) as
\begin{align} \label{infsd}
I(X,Y:\Lambda) = H(X,Y) - \sum_\lambda p(\lambda) H_\lambda(X,Y) = H(X,Y) .
\end{align}
Thus, for $p(x,y)=1/4$ as per Eq.~(\ref{uniform}), the minimum mutual information required to achieve any value $S$ of the CHSH parameter is  given by
\beq \label{isd}
I_{SD}(S) \coloneqq 2~{\rm bits}
\eeq
for superdeterministic models. It follows that such models have the highest possible degree of measurement dependence (since $I(X,Y:\Lambda)\leq H(X,Y)\leq 2$ for any model of the CHSH scenario).

More generally, it is straightforward to write down a superdeterministic and separable model for any prior distribution $p(x,y)$ and any set of statistical correlations $\{p(a,b|x,y)\}$, which may or may not satisfy the no-signaling condition~(\ref{nonsig}), by generalising the outcome deterministic model in~Eq.~(11.24) of Ref.~\cite{hallbrans} to the superdeterministic case. In particular, define $\lambda\coloneqq (\alpha,\beta,\xi,\zeta)$, where $\alpha,\beta,\xi,\zeta$ range over the possible values of $a,b,x,y$, respectively, and define
\begin{align}
&p(\lambda)\coloneqq  p(a{=}\alpha,b{=}\beta|x{=}\xi,y{=}\zeta)\,p(x{=}\xi,y{=}\zeta), \nn \\
&p(a,b,x,y|\lambda) \coloneqq  \delta_{a,\alpha}\,\delta_{b,\beta}\,\delta_{x,\xi}\,\delta_{y,\zeta} 
\end{align}
(where $\delta$ denotes the Kronecker or Dirac delta as appropriate). The desired correlations are then easily recovered via $p(x,y) = \sum_{\lambda,a,b} p(\lambda) p(a,b,x,y|\lambda)$ and $p(a,b|x,y) = \sum_\lambda p(\lambda) p(a,b,x,y|\lambda) / p(x,y)$.

Finally, note that the type of superdeterminism we considered here might more precisely be referred to as `causal superdeterminism', since all causal influences are from the past to the future, as in Fig.~\ref{fig:causal}.  One could also consider, for example, `retrocausal superdeterminism', in which the past source variable $\Lambda$ is fully determined by the future  measurement selections $X$ and $Y$. However, we will not pursue this possibility here.

\section{Biased measurement choices in the CHSH scenario}
\label{sec:biased}

The analysis of the minimum information costs of measurement dependence in the CHSH scenario, for different causal structures, was restricted in Sec.~\ref{sec:chsh} to the case of unbiased measurement choices, i.e., to $p(x,y)=\frac14$ as per Eq.~(\ref{uniform}). However, one may also consider different distributions for $X$ and $Y$. A question then immediately arises: if Alice and Bob are allowed to make any choice of $p(x,y)$, how does the information cost behave? 

A simple example suggests that this cost will go down, i.e, that unbiased measurement choices correspond to the worst-case scenario. In particular, starting with the form of the causal model given in Table~\ref{tab2}, define a new model by replacing each of $p$ and $\tilde p$ by 0 and the column for $p(\lambda)$ by the distribution $\{q^2,q(1{-}q),q(1{-}q),(1-q)^2\}$, for some $0<q<1$. This yields a causal model having the maximum possible value of the CHSH parameter, $S=4$ (note that we cannot take $q$ strictly equal to 0 or 1, as otherwise not all settings are possible and one cannot calculate $S$). Further, by construction, this model has $p(x=0)=p(y=0)=q$ for the distributions of Alice and Bob's measurement choices, and a mutual information $I(X,Y:\Lambda)=H(X,Y) =H(\Lambda)=2h(q)$. Thus, as $p(x)$ and $p(y)$ become more biased, i.e., as  $q$ approaches 0 or 1, the mutual information required for this model becomes arbitrarily small, while still maximally violating the CHSH inequality.

We will show here, more generally, that unbiased measurement choices do indeed correspond to the worst-case scenario: for any factorisable distribution $p(x,y)=p(x)p(y)$ as per Eq.~(\ref{fact}),  the minimal informations calculated in Sec.~\ref{sec:chsh} are sufficient to achieve any given value of the CHSH parameter, for each of the causal structures considered~\cite{footfact}. We also obtain upper bounds on the amount of mutual information required for arbitrary $p(x)$ and $p(y)$, which allow us to significantly strengthen the above example by showing that the minimal mutual information required can always be made arbitrarily close to zero  in the limit of highly biased choices, for any value of $S$.

\subsection{Less mutual information is needed for\\  biased choices}

To demonstrate the results above, we construct explicit models, for any $p(x,y)=p(x)p(y)$, that require no more mutual information to implement than the unbiased models for retrocausal, causal, zigzag, one-sided and superdeterministic measurement dependence in Sec.~\ref{sec:chsh}. These constructions are closely related to the latter models, and make use of an entropic property peculiar to those models.  In particular, we will rely on the following Lemma.
\medskip
\\
{\bf Lemma:} If two measurement dependent models, $M$ and $M'$, with the same range of $X,Y,\Lambda$, satisfy $p_{M^\prime}(\lambda|x,y)=p_M(\lambda|x,y)$ for all $\lambda, x,y$, and the entropy of $p_M(\lambda|x,y)$ is independent of $x$ and $y$, then their mutual informations satisfy
\beq
I_{M'}(X,Y:\Lambda) = I_M(X,Y:\Lambda) + H_{M'}(\Lambda) - H_M(\Lambda),
\eeq
where $H(\Lambda)$ denotes the entropy (in bits) of $\Lambda$.
\medskip
\\
{\it Proof:} Equation~(\ref{infgen}) for mutual information can be rewritten as
$I(X,Y:\Lambda) = H(\Lambda) - \sum_{x,y} p(x,y) H_{xy}(\Lambda)$,
where $H_{xy}(\Lambda)$ denotes the entropy of $p(\lambda|x,y)$. Hence, under the assumptions of the Lemma, one has
\begin{align}
&I_{M'}(X,Y:\Lambda) - I_M(X,Y:\Lambda) = H_{M'}(\Lambda) - H_M(\Lambda) \nn \\ 
&\qquad-\sum_{x,y}\left[ p_{M'}(x,y)H_{M',xy}(\Lambda)- p_M(x,y) H_{M,xy}(\Lambda)\right] \nn \\
&~~= H_{M'}(\Lambda) - H_M(\Lambda) \nn \\ 
&\qquad-\sum_{x,y}\left[ p_{M'}(x,y)- p_M(x,y) \right] H_{M,xy}(\Lambda) \nn \\
&~~= H_{M'}(\Lambda) - H_M(\Lambda) \nn \\ 
&\qquad -H_{M,x_0y_0}(\Lambda)\sum_{x,y}\left[ p_M(x,y)- p_{M'}(x,y) \right] \nn \\
&~~= H_{M'}(\Lambda) - H_M(\Lambda) ,	
\end{align}
as desired, where $x_0$ and $y_0$ are arbitrary values in the ranges of $X$ and $Y$.
$\blacksquare$

\medskip

We first consider the retrocausal model in Table~\ref{tab1}, which has mutual information $I_R(S)$ as per Eq.~(\ref{imin}) for a given value of the Bell parameter $S$. It is convenient to denote the various probabilities and other quantities appearing in this model by the subscript $R$. Thus, for example, $p_R(x,y)=\frac14$. We then construct, for an arbitrary  prior distribution $p_{R'}(x,y)=p_{R'}(x) p_{R'}(y)$, a corresponding model $R'$ defined by $p_{R'}(\lambda|x,y)\coloneqq p_R(\lambda|x,y)$, $A_{R',x}(\lambda)\coloneqq A_{R,x}(\lambda)$, and $B_{R',y}(\lambda)\coloneqq B_{R,y}(\lambda)$. Note that all other properties of $R'$ can be calculated via Bayes theorem, e.g.,  $p_{R'}(\lambda) = \sum_{x,y} p_{R'}(x,y) p_{R'}(\lambda|x,y) = \sum_{x,y} p_{R'}(x) p_{R'}(y) p_R(\lambda|x,y)$. It follows immediately from Eqs.~(\ref{abav}) and~(\ref{chsh}) that these models have the same value of the Bell parameter $S$, i.e., $S = S_{R'}=S_R$ ($=4 - 8 p$ from Table~\ref{tab1}). Further, the conditions of the Lemma are satisfied (with $H_{R,xy}(\Lambda)=h(p)+(1-p)\log_2 3$ for all $x,y$ and $p_R(\lambda)=\frac14$), yielding
\beq \label{rprime}
I_{R'}(X,Y:\Lambda) = I_R(S) + H_{R'}(\Lambda) - 2 \leq I_R(S),
\eeq
noting that the entropy of $p_{R'}(\lambda)$ is bounded above by 2~bits (since $\lambda$ only takes 4 possible values). Thus, as claimed, the minimum mutual information required to implement any given violation of the CHSH inequality via a retrocausal model is never greater than $I_R(S)$ in Eq.~(\ref{imin}), irrespective of the choice of $p(x)$ and $p(y)$.

We proceed similarly for causal and zigzag models. In particular, consider the optimal causal model in Table~\ref{tab2}, and denote all quantities appearing in this model via the subscript $C$. A corresponding model $C'$ with arbitrary prior distribution $p_{C'}(x,y)=p_{C'}(x)p_{C'}(y)$ is then defined via $p_{C'}(\lambda|x,y)\coloneqq p_{C}(\lambda|x,y)$, $A_{C',x}(\lambda)\coloneqq A_{C,x}(\lambda)$, and $B_{C',y}(\lambda)\coloneqq B_{C,y}(\lambda)$. Note that this model is also causal, as it follows via repeated applications of Bayes theorem, and using $p_{C'}(\lambda|x,y)= p_{C}(\lambda|x,y)$ and $p_C(x,y|\lambda)=p_C(x|\lambda) p_C(y|\lambda)$, that
\begin{align}
p_{C'}(x,y|\lambda) &= \frac{p_{C'}(\lambda|x,y)p_{C'}(x)p_{C'}(y)}{p_{C'}(\lambda)}\nn\\
&= \frac{p_C(\lambda)}{p_{C'}(\lambda)}\, \frac{p_C(x|\lambda)p_{C'}(x)}{p_C(x)}\, \frac{p_C(y|\lambda)p_{C'}(y)}{p_C(y)}\nn\\
\label{causaltoo}
&=p_{C'}(x|\lambda) \, p_{C'}(y|\lambda),
\end{align}
in agreement with Eq.~(\ref{causalmd})~\cite{footcaus}. It follows that $S = S_{C'}=S_C$ ($= 4 - 8  p\tilde p$ from Table~\ref{tab2}) and, noting the conditions of the Lemma are satisfied (with $H_{C, xy}(\Lambda)=h(p)+h(\tilde p)$ and $p_C(\lambda)=\frac14$), that
\beq \label{cprime}
I_{C'}(X,Y:\Lambda) = I_C(S) + H_{C'}(\Lambda) -2 \leq I_C(S)
\eeq
for any value of the CHSH parameter $S$. Thus, in analogy to the retrocausal case above, no more than $I_C(S)$ in Eq.~(\ref{icausal}) is required to implement any given violation of the CHSH inequality via causal measurement dependence, irrespective of the choice of $p(x)$ and $p(y)$. A similar result immediately follows for zigzag measurement dependence via the equivalence discussed in Sec.~\ref{sec:zigzag}.

For the case of one-sided causal measurement dependence we again consider the model in Table~\ref{tab2}, but for the choice $\tilde p=\half$ (see Sec.~\ref{sec:one-sided}). Denoting the quantities in this model by the subscript $OS$, a corresponding model $OS'$ with arbitrary prior distribution $p_{OS'}(x,y)=p_{OS'}(x)p_{OS'}(y)$ is then defined via $p_{OS'}(\lambda|x,y)\coloneqq p_{OS}(\lambda|x)$ (implying $OS'$ is also one-sided as per Eq.~(\ref{onesided})), and $A_{OS',x}(\lambda)\coloneqq A_{OS,x}(\lambda)$, $B_{OS',y}(\lambda)\coloneqq B_{OS,y}(\lambda)$. It follows again from Eqs.~(\ref{abav}) and~(\ref{chsh}) that $S = S_{OS'}=S_{OS}$ ($=4-4p$ from Table~\ref{tab2} with $\tilde p=\half$). Further, the conditions of the Lemma are satisfied (with $H_{OS,xy}(\Lambda)=1+h(p)$), yielding
\beq \label{osprime}
I_{OS'}(X,Y:\Lambda) = I_{OS}(S) + H_{OS'}(\Lambda) -2 \leq I_{OS}(S)
\eeq
for any value of the CHSH parameter $S$. Thus, no more than $I_{OS}(S)$ in Eq.~(\ref{ios}) is always sufficient for a one-sided model of Bell nonlocality in the CSHS scenario.

Finally, for superdeterministic models it follows trivially from Eq.~(\ref{infsd}) that any such model, $SD'$ say, with prior distribution $p_{SD'}(x,y)$, requires a mutual information
\beq  \label{sdprime}
I_{SD'}(X,Y:\Lambda) = H_{SD'}(X,Y) \leq 2~{\rm bits}~=I_{SD}(S), 
\eeq
independently of the value of $S$, where the inequality is an immediate consequence of $H(X,Y)\leq 2$ for any joint distribution $p(x,y)$, and $I_{SD}(S)$ is defined in Eq.~(\ref{isd}). Thus, again, the unbiased prior $p(x,y)=\frac14$ is the worst-case scenario.

\subsection{Explicit bounds on mutual information for biased choices}

The left hand sides of Eqs.~(\ref{rprime}) and (\ref{cprime})--(\ref{sdprime}) can be explicitly evaluated for any given  distributions $p(x)$ and $p(y)$, via calculation of the corresponding distribution $p(\lambda)$ and entropy $H(\Lambda)$.  This provides corresponding upper bounds on the mutual informations required to implement a given violation of the CHSH Bell inequality for retrocausal, causal (or zigzag), one-sided and superdeterministic measurement dependence, respectively. We explicitly calculate these bounds here, and show that they approach zero for sufficiently biased $p(x)$ and $p(y)$. 

It is convenient to define the biases of the distributions $p(x)$ and $p(y)$ via 
\beq \label{biases}
\epsilon_X\coloneqq p(x=0)-p(x=1), ~
\epsilon_Y\coloneqq p(y=0)-p(y=1),
\eeq
respectively. Thus, $\epsilon_X,\epsilon_Y\in[-1,1]$ (although for the calculation of $S$ we cannot take $\epsilon_X,\epsilon_Y$ strictly equal to $\pm1$, as otherwise not all settings are possible). Further, for the specific models in Tables~\ref{tab1} and~\ref{tab2} we have $p(\lambda|x,y)=p(x,y|\lambda)$, yielding
\begin{align}
&p_{R}(\lambda_{\mu\nu}|x,y) =\frac{1-p}{3} + \delta_{x,\bar\nu}\,\delta_{y,\bar\mu}\,\frac{4p-1}{3},\nn\\
&p_{C}(\lambda_{\mu\nu}|x,y) = \frac{1+(-1)^{x+\nu}(1-2p)}{2}  \frac{1+(-1)^{y+\mu}(1-2\tilde p)}{2},\nn\\
&p_{OS}(\lambda_{\mu\nu}|x,y) = \frac{1+(-1)^{x+\nu}(1-2p)}{2} 
\end{align}
(recall that the latter is obtained from the preceding case by taking $\tilde p=\half$). The distributions of $\Lambda$ for the models $R'$, $C'$ and $OS'$ can then be calculated, using $p_{M'}(\lambda)= \sum_{x,y} p(x)p(y)p_M(\lambda|x,y)$, as
\begin{align}
p_{R'}(\lambda_{\mu\nu}) & = \frac{1-p}{3} + \frac{1 - (-1)^\nu \epsilon_X}{2}\frac{1 - (-1)^\mu \epsilon_Y}{2} \frac{4p-1}{3}, \nn \\
p_{C'}(\lambda_{\mu\nu}) & = \frac{1+(-1)^\nu \epsilon_X(1-2p)}{2}\frac{1+(-1)^\mu \epsilon_Y(1-2\tilde p)}{2}, \nn \\
p_{OS'}(\lambda_{\mu\nu}) & = \frac{1+(-1)^\nu \epsilon_X(1-2p)}{4},
\end{align}
with corresponding entropies
\begin{align}
H_{R'}(\Lambda) & = H\Big(\Big\{\frac{1-p}{3} + \frac{1\pm \epsilon_X}{2}\,\frac{1\pm \epsilon_Y}{2}\, \frac{4p-1}{3}\Big\}\Big), \nn \\
H_{C'}(\Lambda) & = h\Big(\frac{1+\epsilon_X(1-2p)}{2}\Big) +h\Big(\frac{1+\epsilon_Y(1-2\tilde p)}{2}\Big), \nn \\
H_{OS'}(\Lambda) & = 1 + h\Big(\frac{1+\epsilon_X(1-2p)}{2}\Big) .
\end{align}

Substitution of these entropies into Eqs.~\eqref{rprime}, \eqref{cprime} and~\eqref{osprime} leads to explicit expressions for the mutual informations $I_{R'}(X,Y:\Lambda)$, $I_{C'}(X,Y:\Lambda)$ and $I_{OS'}(X,Y:\Lambda)$, as desired. For example, using Eq.~(\ref{imin}) and recalling that $S=4-8p$ for the retrocausal model in Table~\ref{tab1}, we have
\begin{align} \label{biasr}
I_{R'}(X,Y:\Lambda)&= H\Big(\Big\{\frac{4+S}{24} + \frac{1\pm \epsilon_X}{2}\frac{1\pm \epsilon_Y}{2} \frac{2-S}{6}\Big\}\Big)\nn\\
&\qquad  -h\left( \frac{4-S}{8} \right) - \frac{4+S}{8} \log_2 3 .
\end{align}
Similarly, one finds
\begin{align} \label{biasc}
I_{C'}(X,Y:\Lambda) &=  h\left(\frac{1+\epsilon_X(1-2p)}{2}\right) - h(p) \nn\\
& ~~ + h\left(\frac{1+\epsilon_Y(1-2\tilde p)}{2}\right) - h(\tilde p)
\end{align}
(where $S=4-8p\tilde p$), and
\beq \label{biasos}
I_{OS'}(X,Y:\Lambda) = h\left(\frac{1+\epsilon_X(S/2-1)}{2}\right) - h(S/4) .
\eeq
One also has directly from Eqs.~(\ref{sdprime}) and~(\ref{biases}) that
\beq \label{biassd}
I_{SD'}(X,Y:\Lambda) = h\left(\frac{1+\epsilon_X}{2}\right) +h\left(\frac{1+\epsilon_Y}{2}\right).
\eeq
These mutual informations reduce to $I_R(S), I_C(S)$, $I_{OS}(S)$ and $I_{SD}(S)$ for the unbiased case, $\epsilon_X=\epsilon_Y=0$, as expected. Equations~(\ref{biasr})--(\ref{biassd}) also provide upper bounds for the minimum mutual informations required to implement retrocausal, causal one-sided and superdeterministic models, respectively, for any given biases $\epsilon_X,\epsilon_Y$ and CHSH parameter $S$~\cite{footnonoptimal}. It may be verified that they decrease monotonically to zero as $\epsilon_X, \epsilon_Y\rightarrow\pm1$. Thus, an arbitrarily small amount of mutual information is required in the limit of extreme bias.

\section{General and singlet state models}
\label{sec:singlet}

In Sec.~\ref{sec:chsh} we determined optimal models for Bell nonlocality in the CHSH scenario, under various causal constraints on measurement dependence, and found the ordering
\beq \label{genorder}
I_R(S) < I_C(S) = I_Z(S) < I_{OS}(S) < I_{SD}(S)
\eeq
for the minimum informations required to model any given value $S \in (2,4)$ of the CHSH parameter. Recall that the subscripts denote retrocausal, causal, zigzag, one-sided and superdeterministic measurement dependence, respectively. 

It would be of interest to determine whether an analogous ordering holds more generally, beyond the CHSH scenario. This however appears to be a difficult problem, as optimisation of models becomes harder for increasing numbers of inputs and outputs (including for continuous ranges). In contrast, nevertheless, it is straightforward to demonstrate the result
\beq \label{weakorder}
I_R \leq I_C = I_Z \leq I_{OS} \leq I_{SD} ,
\eeq
for separable models of any given set of joint correlations $\{p(a,b|x,y)\}$, where $I_C$, for example, denotes the minimum mutual information required to generate the set $\{p(a,b|x,y)\}$ under the constraint of causal measurement dependence. This ordering is an immediate logical consequence of the definitions of the corresponding types of measurement dependence in Sec.~\ref{sec:chsh}. In particular, all separable models have retrocausal implementations~\cite{footretro}; causal and zigzag measurement dependence are formally equivalent; one-sided measurement dependence is formally equivalent to a special case of causal measurement dependence; and superdeterministic models have the maximum possible value of mutual information. 

Hence, the source of the difficulty in generalising Eq.~(\ref{genorder}) lies in determining whether a strict ordering obtains. 
In the remainder of this section we consider the case of separable models of spin  measurements on maximally-entangled two-qubit states, which allow at least two possible measurement selections for each observer, and argue that
\beq \label{conj}
I_R^{me} < I_C^{me} \leq I_{OS}^{me} < I^{me}_{SD}
\eeq
for this case. In particular, we give evidence that  the first inequality is strict for this case, and formally prove strictness for the last inequality.

First, note that we can restrict attention to the singlet state, since all other such states differ by local rotations, corresponding to a simple relabelling of the measurement settings (where the mutual information in Eq.~(\ref{infgen}) is invariant under any such relabelling, including for continuous ranges of measurement settings $X,Y$ and of $\Lambda$). For the singlet state, the measurement selections $x$ and $y$ correspond to spin directions on the unit sphere; the measurement results $a,b=\pm1$ correspond to spin `up' and spin `down'; and any separable model must reproduce the correlations 
\beq \label{singlet}
p(a,b|x,y) = \frac{1}{4}(1 - ab \,x\cdot y)
\eeq
for some prior distribution $p(x,y)$.

Second, there is a known separable model of spin correlations for the singlet state~\cite{hall2010}, with a mutual information no greater than $\sim 0.066$ bits for any choice of $p(x,y)$~\cite{relaxed}. Since this model trivially has a retrocausal implementation~\cite{footretro}, it follows that
\beq \label{irme}
I_R^{me} \lesssim 0.066~ {\rm bits}
\eeq
for any $p(x,y)$. For the special case of the CHSH scenario, in which $x$ and $y$ are each restricted to two orthogonal directions, $x=x_0, x_1$ and $y=y_0, y_1$, lying in a common plane with $y_0$ bisecting $x_0$ and $x_1$, and with prior distribution $p(x,y)=1/4$, it can be checked that this model reduces to the model in Table~\ref{tab1} with $p = \frac{1-1/\sqrt{2}}{2}$ and $S=S_Q$~\cite{fried2019}, and has the same mutual information, $I_R(S_Q)\sim 0.046$~bits in Eq.~(\ref{irsq})~\cite{fried2019,relaxed}. It follows via Eq.~(\ref{icgtrir}) that, for the CHSH scenario at least, this singlet-state model can only be implemented retrocausally, suggesting more generally that $I_R^{me} < I_C^{me}$ as per the first inequality in Eq.~(\ref{conj}).

Third, there is a known one-sided separable model of spin correlations for the singlet state, with a mutual information no greater than $\log_2 (2/\sqrt{e})$~bits for any choice of $p(x,y)$~\cite{degorre2005,bg2011,hallbrans}. Thus, using Eq.~(\ref{weakorder}),
\beq \label{icme}
I_C^{me} \leq I_{OS}^{me} \leq  \log_2 \frac{2}{\sqrt{e}} \sim 0.279~{\rm bits}.
\eeq

Fourth, for any superdeterministic model of the singlet state the settings are fully determined by $\lambda$, implying that $p(x,y|\lambda)=\delta_{x,f_X(\lambda)}\,\delta_{y,f_Y(\lambda)}$ for two functions $f_X$ and $f_Y$. Hence, if $p(x,y)$ is supported on a set of nonzero measure, then $H_\lambda(X,Y)=-\infty$ and $H(X,Y)\leq 2 \log_2 (4\pi)$, yielding $I_{SD}^{me} = \infty$
via Eq.~(\ref{infgen}). Alternatively, if $p(x,y)$ is only supported on a discrete set of directions (e.g., as in the CHSH scenario), then $H_\lambda(X,Y)=0$, yielding $I_{SD}^{me}=H(X,Y)$. But for the latter case $I_{OS}\leq H(X)<H(X)+H(Y)=H(X,Y)$ for any one-sided model, where the strict inequality follows since equality can hold only for the trivial case $H(Y)=0$, i.e, only one possible measurement selection for Bob (and we have assumed that the factorisability constraint of Eq.~\eqref{fact} is satisfied). Hence, for either alternative we have 
\beq
I_{OS}^{me} < I_{SD}^{me} 
\eeq
as per the final inequality in Eq.~(\ref{conj}).

It would be interesting to find a separable model of the singlet state with causal measurement dependence and a mutual information lying strictly between the values in Eq.~(\ref{irme}) and~(\ref{icme}).  Such a model would provide evidence that the central inequality in Eq.~(\ref{conj}) is in fact also strict. Finally, note that the above models can be easily generalised to models of spin measurements on noisy singlet states (`Werner states'~\cite{werner}), by mixing them with a model having random outcomes independently of the measurement directions.

\section{Conclusions}
\label{sec:con}

We have determined the minimum mutual information required to implement separable models of Bell nonlocality, under various causal constraints on measurement dependence, for the case of unbiased measurement choices in the CHSH scenario, (Sec.~\ref{sec:chsh}). This leads to a  monotonic ordering of the information-theoretic resources required to implement each of retrocausal, causal, zigzag, one-sided and superdeterministic measurement dependence, as per Eq.~(\ref{genorder}) (see also Fig.~\ref{fig:I_vs_S}). In particular, retrocausal models require strictly less mutual information to implement than causal models, for any given violation of the CHSH Bell inequality. A similar result holds for models of maximally entangled two-qubit states (Sec.~\ref{sec:singlet}). It follows that some measurement dependent models in the literature have no causal implementation~\cite{hall2010,fried2019}.

The underlying reason for why causal measurement dependence, as per Fig.~\ref{fig:causal}, inherently requires more mutual information than retrocausal measurement dependence, as per Fig.~\ref{fig:retro}, is the independence condition~(\ref{causalmd}) for the former case. In general, the more independence conditions imposed by a causal structure~\cite{pearl}, the higher the required mutual information is expected to be.

It is worth noting that while there is no known means of implementing retrocausal models as in Fig.~\ref{fig:retro} (with the depicted timelike and spacelike separations), the result that such models require strictly less resources than causal models is of some theoretical interest in itself. In particular, the relative efficiency of such models for simulating Bell nonlocality provides a further argument for retrocausality, in addition to existing arguments in the literature~\cite{price1,price2,aharonov,leifer,argaman}. 

We have also constructed corresponding optimal models  for each of the considered causal constraints, which have fully deterministic measurement outcomes and no superluminal signaling (Sec.~\ref{sec:chsh}). Thus, an implementation of any of these models subverts the security of device independent information protocols based on Bell nonlocality in the CHSH scenario, with an adversary able to in principle have full knowledge of a cryptographic key or sequence of random numbers generated by such a protocol (see also Sec.~\ref{sec:intro}). This is of particular interest for the case of causal measurement dependence, for which the corresponding models can be easily implemented in practice by an adversarial device manufacturer (see Sec.~\ref{sec:causal}) and which require at most $I_C(S_Q)\sim 0.080$ bits of mutual information to achieve the maximum quantum violation of the CHSH inequality, as per Eqs.~(\ref{icsq}) and~(\ref{cprime}). 

We have further constructed explicit separable models that achieve violation of the CHSH Bell inequality by any specified amount, for any biased distribution of measurement settings $p(x,y)=p(x) p(y)$, and used these models to show that the case of unbiased settings, $p(x,y)=\frac14$, has the highest information cost irrespective of the causal structure, and that the information cost approaches zero in the case of extreme bias (Sec.~\ref{sec:biased}). It would be of interest to improve on the  upper bounds given by $I_{R'}(X,Y:\Lambda)$, $I_{C'}(X,Y:\Lambda)$ and $I_{OS'}(X,Y:\Lambda)$ in Eqs.~(\ref{biasr})--(\ref{biasos}), by calculating the minimum possible mutual information required for a given choice of $p(x)$ and $p(y)$.

We note that the above models can easily be extended to include some further measurement choices on either side (or both), whose corresponding outputs are perfectly correlated (or anticorrelated) with some other outputs---e.g., to add $x=2$ with $A_2(\lambda)\coloneqq B_0(\lambda)$. Such correlations are useful in establishing cryptographic keys~\cite{bellreview,kofler2006,acin06}, and it would be of interest to calculate the corresponding mutual information requirements for such extended models, under various causal constraints and for relevant distributions of settings $p(x,y)$.

It would also be of significant interest, as noted in Sec.~\ref{sec:singlet}, to determine whether there is a causal separable model for the spin correlations of maximally entangled two-qubit states that requires strictly less mutual information than any one-sided separable model.

Finally, note that the various causal structures considered in this paper (Figs.~\ref{fig:causvsretro}, \ref{fig:one-sided}, \ref{fig:zigzag} and \ref{fig:zigzag2}) differ only in the causal relations between $X, Y$ and $\Lambda$: in all cases $A$ and $B$ are directly influenced by $X$ and $\Lambda$, and by $Y$ and $\Lambda$, respectively. A possible direction for future work would be to also consider structures with different causal relations involving $A$ and $B$---e.g., to also consider retrocausal influences from $A$ and $B$ to $X, Y$ and/or $\Lambda$.

\acknowledgments
We thank N Gisin for early discussions. MH is grateful to V Vijendran for discussions and for raising the question of zigzag causality.

\appendix
\section{Properties of $I_1(S)$ and $I_2(S)$}
\label{app}

Here we give a partial formal proof of some properties of the curves $I_1(S)$ and $I_2(S)$ depicted in Fig.~\ref{fig:I1_I2}. In particular, we show that these curves have a common tangent at the intersection point $S=S_0$, that $I_1(S)$ is convex, and that convexity of $I_2(S)$ corresponds to the monotonicity of $f(p)/(4-S)$ with respect to $S$, which can be verified numerically.

Note first from Eqs.~(\ref{bnd_S_causal}) and~(\ref{ilam}) that
\begin{align} \label{i1prime}
	I'_1(S) &= \frac{(\textup{d}/\textup{d}p)[2 - 2h(p)]}{(\textup{d}/\textup{d}p)[4-8p^2]} = \frac{h'(p)}{8p},
\end{align}
and that, using the identities $f(p)=p \, h'(p)$ and $f(p)=f(p^*)$ (and considering $p^*$ to be a function of $p$),
\begin{align}
	I'_2(S) &= \frac{(\textup{d}/\textup{d}p)[2 - h(p) - h(p^*)]}{(\textup{d}/\textup{d}p)[4-8pp^*]}\nn\\
	&=\frac{h'(p)+p^{*'}h'(p^*)}{8(p^*+pp^{*'})}\nn\\
	&= \frac{f(p)/p+p^{*'}f(p^*)/p^*}{8(p^*+pp^{*'})}\nn\\
	&=\frac{f(p)}{8pp^*} = \frac{h'(p)}{8p^*} .
	\label{i2prime}
\end{align}
Hence, it directly follows for $p=p^*=p_0$ that $I'_1(S_0) = I'_2(S_0) = \frac{h'(p_0)}{8p_0} \sim 1.059$, i.e., that the curves $I_1(S)$ and $I_2(S)$ in Fig.~\ref{fig:I1_I2} have a common tangent at $S=S_0$. Note the right hand sides of Eqs.~(\ref{i1prime}) and~(\ref{i2prime}) are positive, verifying that $I_1(S)$ and $I_2(S)$, and accordingly $I_C(S)$ in Eq.~(\ref{icausal}), are monotonically increasing with $S$.

Via a similar calculation as in Eq.~\eqref{i1prime}, one further finds that 
\begin{align}
I_1''(S) &= \frac{1}{128p^3} \left( \log_2\frac{1-p}{p} + \frac{1}{(1-p)\log_e 2} \right) > 0 
\end{align}
(for $0< p\leq \half$), which shows that $I_1(S)$ is convex.

Finally, using $S=4-8pp^*$ for $S\geq S_0$, it follows from Eq.~(\ref{i2prime}) that the curve $I_2(S)$ is convex for $S\geq S_0$ if and only if $f(p)/(4-S)$ is a monotonic increasing function of $S$ (where $p$ is obtained from $S$ as the solution to $4 - 8 p p^* = S$, which in general has to be found numerically). This can be verified via a numerical plot. Alternatively, calculating $I_2''(S)$ in a similar way to Eq.~(\ref{i2prime}) one finds
\begin{align}
	I''_2(S) &= -\frac{(\textup{d}/\textup{d}p)\big[\frac{f(p)}{pp^*}\big]}{64(\textup{d}/\textup{d}p)[pp^*]} \nn\\
	&=\frac{1}{64(pp^*)^2}\frac{(p^*+pp^{*'})f(p)-pp^*f'(p)}{p^*+pp^{*'}},
\end{align}
which can be numerically verified to be positive (one finds that both numerator and denominator in the last fraction above are positive for $p<p_0$, and  negative for $p>p_0$).


\begin{thebibliography}{99}
\bibitem{footentmax} More explicitly, if one writes $p(x,y|\lambda) = \{p,(1{-}p)q_1,(1{-}p)q_2,(1{-}p)q_3\}$, with $p\coloneqq p_{\min}$, $q_i \ge 0$,  and $q_1+q_2+q_3 = 1$, then one has $H_\lambda(X,Y)=h(p)+(1{-}p)H(\{q_1,q_2,q_3\})\leq h(p)+(1{-}p)\log_2 3$ (where $H(\{\cdot\})$ denotes the Shannon entropy of the distribution with components in $\{\cdot\}$). This bound is saturated for $q_1=q_2=q_3=1/3$, and increases monotonically with $p\in[0,1/4]$.
\bibitem{footretro} More generally, a retrocausal model can be constructed for any given set of probabilities $\{p(\lambda|x,y), p(a|x,\lambda), p(b|y,\lambda)\}$ that generates separable correlations $\{p(a,b|x,y)\}$ as per Eqs.~(\ref{bayes}) and~(\ref{sep}), even for the case of an arbitrary given prior distribution $p(x,y)$. In particular, one adds a new random variable, $\Lambda_0 = \lambda_0$ say, that causally influences $X$ and $Y$ in Fig.~\ref{fig:retro} but is not directly connected to any other variable, such that $p(x,y)=\sum_{\lambda_0} p(\lambda_0)p(x|\lambda_0)p(y|\lambda_0)$ (this corresponds to generating the desired $p(x,y)$ via shared randomness). The measurement choices $x$ and $y$ obtained via $\Lambda_0$ then retrocausally influence $\Lambda$, to generate the desired distribution $p(\lambda|x,y)$, which in turn causally influences $A$ and $B$, to generate the desired outcome probabilities $p(a|x,\lambda), p(b|y,\lambda)$. Note that such a retrocausal implementation of measurement dependence generates a corresponding distribution $p(\lambda)=\sum_{x,y}p(\lambda|x,y)p(x,y)$ for $\Lambda$, and that the measurement dependence, as quantified by $I(X,Y:\Lambda)$, is unchanged by the additional variable $\Lambda_0$ (since $X$ and $Y$ `screen off' $\Lambda_0$ from $\Lambda$).
\bibitem{footfact} The factorisability assumption for $p(x,y)$ can be dropped here for the cases of retrocausal and superdeterministic measurement dependence.
\bibitem{footcaus} The last line in Eq.~(\ref{causaltoo}) follows via rewriting the second line as $p_{C'}(x,y|\lambda)={u}(\lambda)v(x,\lambda){w}(y,\lambda)$, so that summation over $y$ and $x$ gives $p_{C'}(x|\lambda)={u}(\lambda){v}(x,\lambda){W}(\lambda)$, $p_{C'}(y|\lambda)={u}(\lambda){V}(\lambda){w}(y,\lambda)$ and $1={u}(\lambda){V}(\lambda){W}(\lambda)$, where ${V}$ and ${W}$ denote the summation of ${v}$ and ${w}$ over $x$ and $y$, respectively.
\bibitem{footnonoptimal} The upper bound given by Eq.~(\ref{biassd}) is in fact tight, noting Eq.~(\ref{infsd}), whereas the upper bounds given by Eqs.~(\ref{biasr})--(\ref{biasos}) are not. In particular, one finds that $I_{R'}(X,Y:\Lambda)$, $I_{C'}(X,Y:\Lambda)$ and $I_{OS'}(X,Y:\Lambda)$ do not satisfy the ordering of Eq.~\eqref{weakorder} (which should hold for optimal models) for  all values of $S, \epsilon_X, \epsilon_Y$. (E.g., for $S = 4$, $\epsilon_X \to \pm1$ and $\epsilon_Y = 0$, one has $I_{R'}(X,Y:\Lambda) \to  1/3$ while $I_{OS'}(X,Y:\Lambda) \to 0$.)
\end{thebibliography}
\end{document}